\documentclass[pra,floatfix,showpacs]{revtex4}
\usepackage{graphicx}
\usepackage{dcolumn}
\usepackage{bm}
\usepackage{epsf}
\def\chin{\varphi}

\def\Un{{U}}

\def\sigman{\delta}
\def\lfrac#1#2{{#1/#2}}

\newcommand{\sfrac}[2]{\raisebox{0.095ex}{\scriptsize${\frac{#1}{#2}}$}}

\usepackage{ulem}
\usepackage{cancel}
\usepackage{color}

\newcommand{\sbf}[1]{\mbox{{\scriptsize$\bf{#1}$}}}
\def\ast{*}

\newcommand{\edc}{\end{document}}
\newcommand{\bb} {\begin{thebibliography}}
\newcommand{\eb}{\end{thebibliography}}
\newcommand{\bi}[1]{\bibitem{#1}}
\newcommand{\bc}{\begin{center}}
\newcommand{\ec}{\end{center}}

\newcommand{\be}{\begin{equation}\small}
\newcommand{\ee}{\end{equation}\normalsize}
\newcommand{\bea}{\begin{eqnarray}}
\newcommand{\eea}{\end{eqnarray}}
\newcommand{\ba}{\begin{array}{l}   }
\newcommand{\lab}[1]{\label{#1}}
\newcommand{\ea}{\end{array}}

\newcommand{\dsfrac}{\displaystyle\frac}
\newcommand{\ds} {\displaystyle}
\newcommand{\summa}{\ds\sum}
\newcommand{\dssum}{\summa}
\newcommand{\re}[1]{(\ref{#1})}

\newcommand{\ci}{\cite}

\newcommand{\dsint}{\ds\int}

\def\bfr{{\bf r}}

\def\bfq{{\bf q}}
\def\bfx{{\bf x}}

\newcommand{{\vergul}}{  ,}

\newcommand{\inttau}{\dsint_{0}^{\beta}}

\begin{document}
\title{Phase Transitions in Three-Dimensional Bosonic Systems in Optical Lattices
}
\author{H. Kleinert$^{a}$}\email{h.k@fu-berlin.de}
\author{Z. Narzikulov$^{b}$}\email{narzikulov@inp.uz}
\author{Abdulla Rakhimov$^{a,b}$}\email{rakhimovabd@yandex.ru}
\affiliation{
$^a$ Institut f$\ddot{u}$r Theoretische Physik, Freie Universit$\ddot{a}$t Berlin, Arnimallee 14, D-14195 Berlin, Germany\\
$^b$Institute of Nuclear Physics, Tashkent 100214, Uzbekistan
}

\begin{abstract}

We formulate the Collective Quantum Field Theory
for  three-dimensional bosonic  optical lattices
and evaluate its consequences
in a mean-field approximation to  two collective fields,
proposed by Cooper  et al. \cite{pra83}, and in a lowest-order
Variational Perturbation Theory (VPT).
  It is shown that present
mean-field approximation
predicts some essential features of the experimentally observed
dependence of the critical temperature
on the coupling strength and  a  second -  order quantum phase transition.
In contrast to a recent prediction for atomic gases by Cooper
et. al., we find
no superfluid state with zero condensate fraction.


\end{abstract}
\maketitle
\section{Introduction}

 Optical lattices are
gases of ultracold atoms trapped
 in periodic potentials created by periodically
arranged intersecting
standing waves of laser light. The interest
in experimental
 and theoretical investigations of these artificial crystals
 is caused by the two following factors \ci{Morchrev}:

1) Neutral atoms in these  optical lattices have several
of
attractive features that make them interesting candidates for the
realization of a quantum computer \ci{rous2003}.

2) They may be used to simulate
various lattice models of fundamental
 importance in condensed matter physics.
Since they permit studying in a controlled way
solid-state physics, in which one can  fine-tune
   the interaction strength for various geometries of the lattices.
In particular, it is possible to control the Hamiltonian parameters
and study various regimes of system parameter.

 The lattice of bosons with short - range repulsive pair interaction
 trapped in an optical lattice may be described
by
a Hamiltonian
 of  Bose-Hubbard type:
\begin{equation}
H=-J\sum _{\langle{\sbf i},{\sbf j}\rangle}\hat{b}_{\sbf{i}}^{\dag} \hat{b}_{\sbf{j}}  +\frac{U}{2} \sum
_{\sbf{i}}^{N_s}\hat{b}_{\sbf{i}}^{\dag} \hat{b}_{\sbf{i}}^{\dag} \hat{b}_{\sbf{i}} \hat{b}_{\sbf{i}}
+\sum _{\sbf{i}}^{N_s}(\varepsilon _{\sbf{i}} -\mu )\hat{b}_{\sbf{i}}^{\dag} \hat{b}_{\sbf{i}},
\label{H11}
\end{equation}
where $\hat{b_{\sbf{i}}}^{\dag}$ and $\hat{b_{\sbf{i}}}$ are the bosonic creation
 and annihilation operators on the site $i$; the  sum over $\langle{\sbf i},{\sbf j}\rangle$
 includes only pairs of nearest neighbors;
  $J$ is the hopping amplitude, which is responsible for the
   tunneling of an atom from one site to another neighboring site;
    $U$ is the on site repulsion energy, and $N_s$ the number of sites.

At zero temperature with an  {\it  integer filling factor} $\nu\equiv
N/N_s$, where $N$ is the total number of atoms,
   a system of bosons described by the Hamiltonian
  \re{H11} could be on superfluid (SF) or
  in Mott insulator (MI) phase. Clearly the quantum phase transition (QPT)
  between these two phases is allocated by the dimensionless
interaction strength
parameter $u=U/J$.
 For small  $u$, the hopping term dominates the system,
so that it prefers to be in the SF phase. For large $u \gg1$, on
the other hands,  the system exhibits a MI phase.

A critical interaction  strength $u_{\rm crit}=29.34$  was found for
$d=3$
 by Monte Carlo calculations \ci{svistun}
  at a filling factor $\nu=1$, and agrees
well with the experimental data \ci{trotzky}.

To make for easier reading, we summarize some specific features of
these two phases.
 The SF phase is characterized by a long-range
correlation, a continuous (gapless) excitation spectrum and a finite
compressibility. Since there exists a condensate with a finite
number of particles, $n_{\sbf 0}$, the gauge symmetry is
spontaneously broken in accordance with theorems by  Bogoliubov and
Ginibre. In contrast, in the MI phase, there is no long-range
correlation or breaking of gauge symmetry. The excitation spectrum
has a gap and the system is incompressible, since there is a fixed
number of atoms per-site. The mobility of atoms is completely
different in the two phases. In the SF phase they  can easily move
from one site to another site by tunneling, whereas in the MI phase,
they are localized.

Finite-temperature phases of optical lattices have been studied by quantum
Monte Carlo (QMC) calculations as well as experimentally for $d=3$. As
expected, the system behaves as a normal fluid (NF) at $T>T_c$. A
most interesting observation was made in Refs. \ci{svistun,trotzky}:
In contrast to the system of dilute Bose gases, the critical
temperature is downshifted at the transition to the MI phase.

{Theoretical approaches  based on the Bose-Hubbard
model, which is not exactly  soluble even  in one dimension, have
been summarized recently   in textbooks \ci{lewenstein, ueda,
stoofbook}. Most  of them use  perturbative expansions in powers of
$J/U$ and give qualitatively a good description of phase  transition
boundary
 \ci{Pelster,freericks}. As  to the nonperturbative approaches, they  mainly  exploit
 the Gutzwiller ansatz,  where the wave function is  expanded in local  Fock  states with
   variational   coefficients. Although  such  an approach is good even in   description of
   the dynamics of the system
   \ci{dutta,luhman,amico,vezzani}, since it is exact for $d\rightarrow\infty$,
its reliability decreases  dramatically for  $d=1$.}

 Among various types of the existing mean -  field theories in the
 literature the bosonic dynamical mean-field theory (B-DMFT) seems
 to be most powerful. Being originally proposed by Byczuk and Vollhardt \ci{BV} and further developed by Anders et al.
 \ci{anders} the B-DMFT maps the Bose-Hubbard model onto the self - consistent solution of a bosonic impurity
 model with coupling to a reservoir of normal and condensed bosons. The net output of this procedure is
 delightful. It gives as an accurate description of the phase diagrams, the condensate
 order parameter and other observables of the cubic lattice Bose-Hubbard model as it
 was obtained by QMC calculations. However, although the B-DMFT  is numerically exact and flexible,
 it is computationally expensive, since one has to use continuous time
 QMC evaluations in order to solve its equations. Moreover, strictly speaking, the Hugenholtz - Pines theorem
  (see Subection  IIB below) does not hold in B-DMFT (see Fig. 10 of Ref. \ci{anders}).

The application of non-perturbative renormalization group theory has revealed new scaling properties
  of optical lattices. Rancon and Dupuis \ci{dupius} have recently shown that  thermodynamic quantities of the Bose-Hubbard model
can be expressed using universal scaling functions of the dilute
Bose gas universality class.

 {As to  the Bogoliubov theory, it  provides an  accurate  description of the excitation spectrum
   for  the SF phase,  but fails to describe SF $\rightarrow$ MI transition.  In fact,
  the first application of a
mean-field approach was made in the Hartree-Fock-Popov (HFP)
  approximation to optical lattices by Stoof et. al. \ci{Stoofmakola}.
By studying the dependence of the condensate number $n_{\sbf 0}$ on
$u$, i.e. $n_{\sbf 0}(U/J)$ they observed that $n_{\sbf 0}$ never
reaches zero, even in the strong-coupling limit ($u\rightarrow
\infty$), implying that this approximation is unable to predict a
QPT of SF $\rightarrow $ MI. In contrast to this, the two-loop
approximation by the present authors in \ci{ourknr} suggests the
existence of such a QPT, but the critical value of $u_{\rm crit}$
was found to be rather small: $u_{\rm crit}$(two-loop)$\approx 6$
for $d=3$. So, the question about the power of
 an approximation, based on mean-field theory, other than B-DMFT, to adequately describe  phase diagrams of optical lattices
   remains still open.  It  is, therefore,  desirable to develop a nonperturbative
  approach  which  would be suitable for dimensions  $d=1,2,3$.}

  An alternative approach
to the treatment of dilute Bose gases
 has recently been
proposed
by Fred Cooper et al. \ci{pra83,pra84}
under the name of
{\it leading-order auxiliary field theory} (LOAF)\label{LOA}.
They found a
way of fixing the degeneracy in the
elimination of the interaction
by auxiliary collective pair and density fields
by choosing a special form of a generalized
 Hubbard-Stratonovitch transformation.
  Although their approach gives no QPT for a homogenous Bose gas at zero temperature, it predicts a desirable
  second order BEC transition at finite temperatures
and exhibits  a positive shift
  in the critical temperature $T_c$ that is consistent with Monte Carlo
an other calculations \ci{baym,Kl5l}.
  One of the novel features of that calculations is that
for
  $T>T_c$
 it predicts a novel type of
superfluid phase that does not have a condensate  \ci{pra86}.
Although such a phase has not been observed yet, it
was justified by the existence of a
  nonzero anomalous density $\delta$, in the region $T_c < T \leq T^\ast$, where $T^\ast$ is the transition temperature
  to the normal phase.

 In the present work we shall formulate
 a similar two-collective quantum field theory
 for discrete systems such as
optical lattices and ask the following questions
 \begin{itemize}
    \item
    Does it predict a SF $\rightarrow$ MI quantum phase transition?
    \item
    Does it predict the suppression of $T_{c}$ at large $u$?
    \item
   Does it predict a new phase, mentioned above, for optical lattices either?
\end{itemize}

Our results will be compared with those of another well-known mean-field
approximation, the Hartree - Fock - Bogoliubov
  (HFB) approximation,
which is widely used to describe BEC in homogeneous Bose gases and
in triplons
  \ci{ourtriplon, ourdisorder} in magnetic insulators, and will also be
extended here to optical lattices. Below we use $\hbar=k_{B}=1$.

  The paper is organized as follows.
  In Sections II and III we shall derive   Collective Quantum Field Theory
and
HFB approaches for optical lattices, respectively.
  The results and discussions will be presented in Section IV, and the conclusions will be stated in
  Section V.

\section{Collective Quantum Field Theory of 3D  Bose-Hubbard model}

 In the Wannier  representation the Euclidian action, corresponding to the
Bose-Hubbard
Hamiltonian is given by \ci{ourknr}
  \begin{eqnarray}
 {\cal A} (\psi^{\ast},\psi)&=&\int_{0}^{\beta} d\tau\Bigg\{
\sum_{{\sbf i}}\psi^{\ast}(\bfx_{{\sbf i}}, \tau)[\partial_{\tau}-\mu]
  \psi(\bfx_{{\sbf i}}, \tau)\nonumber\\&-&J\sum_{\langle i,j\rangle}\psi^{\ast}(\bfx_{{\sbf i}}, \tau)\psi(\bfx_{j}, \tau)\nonumber\\
   &+&\frac{U}{2}\sum_{{\sbf i}}\psi^{\ast}(\bfx_{{\sbf i}}, \tau)\psi^{\ast}(\bfx_{{\sbf i}}, \tau)
    \psi(\bfx_{{\sbf i}}, \tau)\psi(\bfx_{{\sbf i}}, \tau)\Bigg\},
    \label{1.1}
     \end{eqnarray}
     where $\mu$
 is the chemical potential and $\beta=1/T$. The lattice points
lie at  the positions \cite{GFCM}
\begin{eqnarray}
{\bf x}_{\sbf i}={\bf i}\,a,
\label{@}\end{eqnarray}
where $a$ is the lattice spacing, and
\begin{eqnarray}
{\bf i}\equiv
(i_1,i_2,\dots,i_d),
\label{1.1a}\end{eqnarray}
are integer-valued vectors.

The partition function $Z$, and the grand thermodynamic potential $\Omega$, can be found as:
\begin{eqnarray}
 Z&=&\dsint{ {D}}\psi^{\ast}{ {D}}\psi e^{-{\cal A} (\psi^{\ast},\psi)},\label{1.8}\\
  { {\Omega}}&=&-{T}\ln{Z}\label{1.2}.
   \end{eqnarray}
 The ground state expectation   value of an operator $\hat{O}(\psi^{\ast},\psi)$
 can be expressed as a functional integral:
\begin{eqnarray}
 \langle \hat{O}\rangle=\frac{1}{Z}\int{\cal{D}}\psi^{\ast}{\cal{D}}\psi
  \hat{O}(\psi^{\ast},\psi)e^{-{\cal A} (\psi^{\ast},\psi)}\label{1.3}.
   \end{eqnarray}
With the help of a
 Hubbard-Stratonovich transformation,
the interaction term in
 \re{1.1} can be eliminated by adding to the
action in the exponent of (\ref{1.8}) a dummy
action
\ci{CQFp}:
 \be
   \ba
   {\cal A} _{\rm pair}[
\psi^\ast,\psi, \Delta,\Delta^\ast]=\dsint_{0}^{\beta} d\tau   \sum_{{\sbf i}} \Bigg\{
        \dsfrac{1}{2U}\Big|\Delta (\bfx_{{\sbf i}}, \tau)-U
 \psi(\bfx_{{\sbf i}}, \tau)\psi(\bfx_{{\sbf i}}, \tau)   \Big|^2
     \Bigg\}.
     \lab{1.4a}
     \ea
     \ee
containing a {\it pair field} $\Delta$.
After this we form the path integral
$\int {\cal D}\Delta {\cal D}\Delta^\ast e^{-{ {\cal A} _{\rm pair}}[
\psi^\ast,\psi, \Delta,\Delta^\ast]}$,
and
 integrate out the pair field.
This
produces a
multiplication
of the partition function $Z$ by a trivial constant factor.

It has been emphasized in \ci{CQFp} and the textbook  \ci{CQFb} that
this procedure  is highly degenerate. Actually, instead of~(\ref{1.4a}), one could just as well have introduced a plasmon field
$\varphi({\bf x},\tau)$
by adding to
 the
action in the exponent of (\ref{1.8}) a dummy action
  \be \ba
   {\cal A} _{\rm pl}[
\psi^\ast,\psi,  \varphi]=\dsint_{0}^{\beta} d\tau   \sum_{{\sbf i}} \Bigg\{-
  \dsfrac{1}{2U}\left[\varphi(\bfx_{{\sbf i}}, \tau) -  U \psi^{\ast}(\bfx_{{\sbf i}}, \tau) \psi(\bfx_{{\sbf i}}, \tau)\right]^2 \Bigg\}
    , \lab{1.4b}
     \ea
     \ee
and forming
a functional integral
integral
$\int {\cal D}\varphi e^{-{ {\cal A} _{\rm pair}}[
\psi^\ast,\psi,\varphi]}$,
which again multiplies $Z$ by a trivial constant.

Diagrammatically, the degeneracy is caused by the fact that the sum
of all collective field diagrams will always produce the same result
if evaluated to {\it all} orders in perturbation theory. Each of
these collective fields reproduces all effects of the interaction if
it is integrated functionally. A difference appears, if the
evaluation is restricted to a mean-field approximation. Then it
depends on the dominance of certain dynamical effects which field is
preferable.

In principle, we can also add a combination of ${\cal A}_ {\rm
pair}$ and ${\cal A} _{\rm pl},$ and still leave the physical
properties of the system unchanged. For instance ${\cal A}_ {\rm
pl}\cosh^2 \theta -{\cal A} _{\rm pair}\sinh^2 \theta $.
Diagrammatically, however, the degeneracy cannot be easily verified
since a calculation of the diagrams to all order is really
impossible. It can only be done to some finite order, for instance
in a loop expansion, so that the mathematical equivalence is
initially of little use.

One method to avoid the degeneracy and
 make the
collective field approach unique has been pointed out a long time
ago  \ci{HEA}. It is based on an extension of the standard effective action
 ${{\Gamma}} [\Psi^\ast,\Psi]$, whose functional expansion terms are
the one-particle irreducible vertex functions of the theory.
The symbol $\Psi$ denotes the expectations of the field $\psi({\it x},\tau)$.
A
unique version of collective fields can be introduced by going to a
 {\it higher effective action} ${{\cal A}} [
\Psi^\ast,\Psi, \Delta,\Delta^\ast, \Phi]$. While the ordinary
effective action ${{\Gamma}} [\Psi^\ast,\Psi]$ is derived from a
 Legendre transformation
of the generating functional of the theory $W[\eta,\eta^\ast]$ in which
additional source terms $\eta\psi^\ast+\eta^\ast\psi$ have been added to
the action, the higher effective action
 is obtained
from the Legendre transformation
of a generating functional $W[\eta,\eta^\ast, j,K,K^\ast]$
in which additional sources have been added to the action
coupled to the density
and the pair fields.
The higher effective action
 will
depend on
the expectations of the fields $\psi,\psi^\ast, \rho\propto\psi^\ast\psi,
\Delta\propto\psi\psi$ and $\Delta^\ast\propto\psi^\ast\psi^\ast$.
 At the end,
it must
merely be {\it extremized}, and no  extra functional integrals 
can cause any double-counting of Feynman diagrams.
The expansion terms in
the higher effective action are the two-particle
irreducible
vertex functions of the theory.

Another method
that also
abandons the fluctuations of the collective fields
in favor of a
collective classical field
has been developed in recent years
from a generalization of a variational approach
to path integrals \ci{FKL} to all orders in perturbation theory.
It was extremely successful and has
led to the most accurate theory of critical phenomena \ci{KS} so far,
named Variational
Perturbation Theory (VPT) (for a review paper see \ci{KLVPT}).

A third method which has  recently been proposed
and applied
\ci{pra83,pra84} uses the combination of both fully fluctuating
collective fields implied by the above dummy action ${\cal A}_ {\rm
pl}\cosh^2 \theta -{\cal A} _{\rm pair}\sinh^2 \theta $ for the
particular value $\sinh\theta=1$. This choice is preferable if we
want the mean-field approximation to exhibit
 excitations that have no energy gap, to comply with
the Nambu-Goldstone theorem. After a trivial change of the
normalization of plasmon and pair fields in the total action ${\cal
A}+{\cal A} _{\rm pl}\cosh^2 \theta-{\cal A} _{\rm
pair}\sinh^2\theta$ one arrives at
\be
  \ba
  {\cal A} ={\cal A} _{\psi}[\psi^{\ast},\psi]+{\cal A} _\varphi [\varphi ]+
{\cal A} _\Delta [\Delta ,\Delta ^{\ast}]\lab{1.5},
  \ea
  \ee
with
\begin{widetext}
\bea\!\!\!\hspace{-3cm}&&\!\!\!\!\!\!
    {\cal A} _\psi[\psi^{\ast},\psi]=\dsint_{0}^{\beta} d\tau\sum_{{\sbf i}}\left\{\psi^{\ast}(\bfx_{{\sbf i}}, \tau)[
  \partial_\tau-\mu+\varphi  (\bfx_{{\sbf i}}, \tau)\cosh\theta
    ]\psi(\bfx_{{\sbf i}}, \tau)\right.\nonumber\\
    \qquad\qquad&&\left.-\sfrac{1}{2}\sinh\theta[\Delta \psi^{\ast}(\bfx_{{\sbf i}}, \tau) \psi^{\ast}(\bfx_{{\sbf i}}, \tau) +\Delta ^{\ast}\psi(\bfx_{{\sbf i}}, \tau) \psi(\bfx_{{\sbf i}}, \tau)]\right\}
  \!  -J\dsint_{0}^{\beta} d\tau\sum_{{\sbf i},{\bf j}}\psi^{\ast}(\bfx_{{\sbf i}}, \tau) \psi(\bfx_{\bf j}, \tau),\lab{1.5}\\&&\!\!\!\!\!\!
       {\cal A} _{\varphi } [\varphi ]=-\dsint_{0}^{\beta} d\tau\sum_{{\sbf i}}\dsfrac{\varphi ^2 (\bfx_{{\sbf i}}, \tau)}{2U}, \quad \quad
         {\cal A} _{\Delta } [\Delta ,\Delta ^{\ast}]=\dsint_{0}^{\beta} d\tau\sum_{{\sbf i}}\dsfrac{\Delta  (\bfx_{{\sbf i}}, \tau)\Delta ^{\ast} (\bfx_{{\sbf i}},
         \tau)}{2U}\lab{aphi}.
  \eea
  \end{widetext}
At the level of for fully fluctuating fields $\varphi$, $\Delta$, $\Delta^\ast$,
 the parameter $\theta $ is still arbitrary,   which   will   be   fixed  in   the  next   section.

Now we consider separately two regions, with and without
a condensed phase.

\subsection{Condensed phase}

In this phase, the $U(1)$ gauge symmetry is spontaneously broken. It
can be studied after a  Bogoliubov   shift of the field
\cite{Yukalovobsor} \be
 \psi(x_{{\sbf i}},\tau)=\psi_{\sbf 0}+\widetilde{\psi}(x_{{\sbf i}},\tau)\lab{6},
  \ee
with
\be
 \psi_{\sbf 0}=\sqrt{\nu n_{\sbf 0}}.\lab{6b}
  \ee
where the  $n_{\sbf 0}=N_{\sbf 0}/N$ is the
 condensate fraction.
It is a constant in the absence of a
 magnetic trap. The fluctuating field $\tilde{\psi}(x,\tau)$ must satisfy the condition:
\be \int_{0}^{\beta} d\tau\sum_{{\sbf i}}\tilde{\psi}(x_{{\sbf i}},\tau)=0.
\ee Substituting (\ref{6}) into (\ref{1.5}), and decomposing the
 quantum field  $\tilde{\psi}({\bf x}_{\sbf i}, t)$ into its real and imaginary parts
 $\psi_{1}({\bf x}_{\sbf i}, t)$ and $\psi_{2}({\bf x}_{\sbf i}, t)$ as
\begin{eqnarray}
 \tilde{\psi}({\bf x}_{\sbf i}, t)=\frac{1}{\sqrt{2}}
  (\psi_{1}({\bf x}_{\sbf i},t)+i\psi_{2}({\bf x}_{\sbf i}, t)),\nonumber\\
  \tilde{\psi}^{\ast}({\bf x}_{\sbf i}, t)=\frac{1}
   {\sqrt{2}}(\psi_{1}({\bf x}_{\sbf i},t)-i\psi_{2}({\bf x}_{\sbf i}, t)),\label{1.12}
   \end{eqnarray}
we may separate the action as follows:
\be
 \ba
  {\cal A} ={\cal A} _{0}+{\cal A}_{2} +{\cal A} _{\Delta}+{\cal A} _{\varphi},\label{@16}
  \ea
  \ee
with
\bea
 \!\!\! \!\!\!\!\!\!\!\!\!\!\!\!\!\!\! {\cal A} _{0}\!\!&=&\!\!-N_{s}\beta\nu n_{\sbf 0}(\mu+Jz_{0})\!
 +\nu n_{\sbf 0}\displaystyle{\sum_{{\sbf i}}}\int_{0}^{\beta} d\tau\left[\cosh\theta\varphi (x_{{\sbf i}},
 \tau)\!\right.\nonumber\\{}&&-\left.\!\sfrac{1}{2}\sinh\theta
  \left(\Delta (x_{{\sbf i}},\tau)\!+\!\Delta ^{\ast}(x_{{\sbf i}},\tau)\right)\right]\!,\\
 \!\!\!\!\!\!\!\!\!\!\!\!  {\cal A} _{2}\!\!&=&\!\!\frac12\displaystyle{\sum_{{\sbf i}}}\int_{0}^{\beta}
  d\tau\sum_{a,b=1,2}\left[ i\varepsilon_{ab}\tilde{\psi}_{a}(x_{{\sbf i}},\tau)\partial_{\tau}\tilde{\psi}_{b}
  (x_{{\sbf i}},\tau)\right.\nonumber\\{}&&+\left.\tilde{\psi}_{a}(x_{{\sbf i}},\tau)X _{a}\tilde{\psi}_{b}
  (x_{{\sbf i}},\tau)\delta_{ab}\right]\nonumber\\
  {}&&-\dsfrac{J}{2}\int_{0}^{\beta} d\tau\displaystyle{\sum_{\langle{\sbf i},{\sbf j}\rangle}}\sum_{a}\tilde{\psi}_{a}(x_{{\sbf i}},\tau)\tilde{\psi}_{a}
  (x_{j},\tau),
   \eea
where ${\cal A} _{\Delta }$ and ${\cal A} _{\varphi }$  are given in
(\ref{aphi}), $\varepsilon_{ab}$ is an antisymmetric tensor with
$\varepsilon_{12}=-\varepsilon_{21}=1, ~ z_{0}=2d$, and
 \be \ba X_{1}=-\mu+\varphi (x_{{\sbf i}},\tau)\cosh\theta-
\frac{1}{2}\sinh\theta\left(\Delta ^{\ast}
 (x_{{\sbf i}},\tau)+\Delta (x_{{\sbf i}},\tau)\right),\\
 \\
 X_{2}=-\mu+\varphi (x_{{\sbf i}},\tau)\cosh\theta+\frac{1}{2}\sinh\theta\left(\Delta ^{\ast}
 (x_{{\sbf i}},\tau)+\Delta (x_{{\sbf i}},\tau)\right)\lab{8}.
  \ea
  \ee
  For a homogenous,
 system the condensate is uniform and it is convenient to decompose the fluctuations into a Fourier series as \ci{HK,danshita}
  \begin{eqnarray}
 \tilde\psi_{a}({\bf x}_{\sbf i}, \tau)=\dsfrac{1}{\beta\sqrt{N_{s}^d}}\dssum_{\sbf q,\omega_{n}}{}'
 \inttau\psi_{a}({\bf q},\omega_n)e^{-i\omega_n \tau}\exp\left[i\bfx_{\bf i}{\bf p}_{\sbf q}\right]\nonumber\\
  \label{9}
   \end{eqnarray}
   where $\omega_{n}=2\pi nT$ are Matsubara frequencies, and
${\bf p}_{\sbf q}\equiv\{q_{1},q_{2},\dots,q_{d}\}~ {2\pi}/{N_sa}$,
with $q_{{\sbf i}}$ running from 1 to $N_{s}-1$
 are the discrete-valued momentum vectors
in the  Brillouin zone. The momentum sum is explicitly
   \be
    \frac{1}{N_{s}}\sum_{\bfq}{}'\equiv\frac{1}{N_{s}^{d}}\sum_{q_{1}=1}
    ^{N_{s}-1}\sum_{q_{2}=1}^{N_{s}-1}\dots\sum_{q_{d}=1}^{N_{s}-1}\lab{10}.
     \ee
The prime on the symbol indicates that the ${\bf p}=0$ -mode is omitted since it is contained
in the subtracted $\psi_{\sbf 0}$. This will be useful
to avoid possible infrared divergencies, especially for $d<3$.

In momentum space, the quadratic term ${\cal A} _{2}$ reads
\be
 {\cal A} _{2}=\frac12\sum_{\bfq,\bfq', m,n}\psi_{a}(\bfq,\omega_{n})G_{ab}^{-1}
  (\bfq,\omega_{n};\bfq',\omega_{m})\psi_{b}(\bfq',\omega_{m})\lab{11},
   \ee
with the propagator
 \begin{eqnarray}
   G(\omega_{n},{\bf q})&=&\frac{1}{\omega_{n}^{2}+{\cal E}^{2}({\bf q})}
    \left(
\begin{array}{cc}
 {\varepsilon}({\bf q})+X_{2}-Jz_0 & \omega_{n}\\
 -\omega_{n} &{\varepsilon}({\bf q})+X_{1}-Jz_0
\end{array}\right)\label{12},
  \end{eqnarray}
  where the bare dispersion ${\varepsilon}({\bf q})$ and phonon dispersion ${\cal E}({\bf q})$ are given by
  \begin{eqnarray}
  {\varepsilon}({\bf q})&=&2J\bigg(d-\summa_{\alpha=1}^{d}\cos (2\pi q_{\alpha}/N_s)\bigg), \label{@22a}
\\
 {\cal E}({\bf q})&=&\sqrt{(X_{1}+{\varepsilon}({\bf q})-Jz_0)
  (X_{2}+{\varepsilon}({\bf q})-Jz_0)}.
   \label{13} \label{@26}
    \end{eqnarray}
In the long-wavelength limit,
$ {\varepsilon}({\bf q})$ behaves like
\begin{eqnarray}
 {\varepsilon}({\bf q})\approx J\frac{4\pi^2}{N_s^2}{\bf q}^2=
J a^{2}{\bf p}^2+\dots~. \label{@}\end{eqnarray} By comparison with
the usual momentum-dependence  of a free single-particle energy
${\bf p}^2/2M$ we identify the particle mass $M=1/2Ja^{2}$.

    Note that in coordinate space the Green function is defined by
 \begin{eqnarray}
 {G}_{ab}{({\bf x}_{\sbf i},\tau;{\bf x}_{\sbf j},\tau')}&\equiv&
   {G}_{ab}{({\bf x}_{\sbf i}-{\bf x}_{\sbf j},\tau-\tau')}\nonumber\\&=&\langle\psi_{a}({\bf x}_{\sbf i},\tau)
   \psi_{b}({\bf x}_{\sbf j},\tau')\rangle\nonumber\\ {}
&=&\frac{1}{N_{s}\beta} \sum_{n}\sum_{q}e^{i\omega_{n}(\tau-\tau')}e^{i\bfq ({\bf x}_{\sbf i}
-{\bf x}_{\sbf j})}\nonumber\\{}&&\times G_{ab}(\omega_{n},\bfq)\label{13_1}.
   \end{eqnarray}
The thermodynamics of the system can be calculated from the
partition function $Z$ functional integral  over all fields
$\psi_{1}, \psi_{2}, \varphi , \Delta $ and $\Delta ^{\ast}$ fields
   \be
   Z=\int{\cal D}\psi_{1}{\cal D}\psi_{2}{\cal D}\varphi {\cal D} \Delta {\cal D}\Delta ^{\ast}e^{-{\cal A} _{0}-{\cal A} _{2}-{\cal A} _{\Delta }-{\cal A} _{\varphi }}\lab{14}.
    \ee
    The first integrations by $\psi_{1}$ and $\psi_{2}$ are Gaussian and may be evaluated easily by using well-known formula
    \bea
 {}&&\dsint{\cal D} \psi_{1}{\cal D} \psi_{2} ~ {\exp\left[-\frac12\sum_{a,b=1,2}
 \int\psi_{a}(x)G_{ab}^{-1}(x,y)\psi_{b}(y) dx dy\right.}\nonumber\\{}&&{-\left.\int j_{1}(x) \psi_{1}(x)dx-\int j_{2}(x)
  \psi_{2}(x)dx\right]}\nonumber\\&=&\sqrt{{\rm Det\,}G}~ \displaystyle{\exp\left[\sum_{a,b=1,2}\int j_{a}(x)G_{ab}(x,y)j_{b}(y)dxdy\right]}\lab{15}.
      \eea
The integrations over the fluctuating collective  fields, however,
cannot be performed exactly, since they are nontrivially contained in $\sqrt{{\rm
Det\,}G}$. As usual in these circumstances, we resort to  the saddle-point approximation
\ci{CQFp,REM1}. In the absence of a trap, we may assume
 the saddle point
to  lie at constant values of
  $\varphi (x_{{\sbf i}},\tau)$ and $\Delta (x_{{\sbf i}},\tau)$:
\be
\ba
 \varphi (x_{{\sbf i}},\tau)=\varphi _{0},\nonumber\\
  \Delta (x_{{\sbf i}},\tau)=\Delta ^{\ast}(x_{{\sbf i}},\tau)=\Delta _{0}\lab{16}.
  \ea
      \ee
Then the integrals over $\psi_{a}$ become trivial and we may use the
formula ${\rm Det}\, G=e^{{\rm Tr}\ln G}$ in Eqs. (\ref{14}) and
(\ref{15}) to derive the following effective potential:
\be
\ba
 \Omega=\displaystyle\frac{T}{2}\displaystyle
 \sum_{q}\displaystyle\sum_{n}\ln(\omega_{n}^{2}+{\cal E}^{2}({\bf q})) + N_{s}\nu n_{\sbf 0} (\varphi '-\Delta)\\ \qquad+\dsfrac{N_{s}\Delta^{2}} {2U\sinh^2\theta}-\dsfrac{N_{s}(\varphi '+\mu+Jz_{0})^2}{2U\cosh^2\theta}\label{17},
 \ea
 \ee
 with
 \be
  \Delta\equiv \Delta_0 \sinh\theta,\quad \varphi '=\varphi _{0} \cosh\theta-\mu-Jz_{0}\label{17_1}.
   \ee
The spectrum of density fluctuations is now
 from (\ref{@26}):
   \be
   {\cal E}^{2}({\bf q})=(\varepsilon({\bf q})+\varphi '-\Delta) (\varepsilon({\bf q})+\varphi '+\Delta) \lab{18}.
   \ee
The sum over $\bf p$ may be calculated in $d=3$ by approximating (\ref{10})
 as follows
\be
\frac{1}{N_{s}}\sum_{\sbf q}f(\varepsilon({\bf q}))\to
\int_{0}^{1}dq_{1}dq_{2}dq_{3}f(\varepsilon_{\sbf q})\label{19}, \ee
with the lattice dispersion:
 \be
 \varepsilon_{\sbf q}=2J\sum_{\alpha=1}^{3}\left[1-\cos\pi q_{\alpha} \right]\lab{20}.
  \ee

  Note that on lattices,
 the momentum  integrals
are always finite so that there is no need for renormalizing the
coupling constant. This is in contrast to  atomic gases. However, if
we want to express the coupling constant in terms of the scattering
length $a_s$ that is observable at low-energy  atomic gases, where
the quadratic coupling constant $g$ must be renormalized to a finite
value $g_R$ by the addition of a diverging integral $1/g_R =1/g+\int
d^3p/(2\pi )^3 \varepsilon({\bf p})$, the relation $a_s=Mg_R/4\pi $
can only be employed only after a corresponding addition of a finite
sum [see the remarks after Eq. (\ref{gasp})].

Another remark concerns the frequency sum in (\ref{17}), which is
initially
divergent. In fact,
 to evaluate a frequency  sum such as $\dssum_{n=-\infty}^{\infty}\ln(a^2+\omega_{n}^{2})$
with $\omega_{n}=2\pi nT$, one must first
differentiate it with respect to
$a$, perform
the summation over $n$, and integrate the result over $a$ \ci{
PI}. This procedure gives an additional divergent constant, which
may be removed by an additive renormalization of the energy
\ci{haugset}. The subtraction can actually be justified by
calculating the path integral as a product of individual integrals
for each slice of a sliced time axis, as introduced originally by
Feynman \ci{PI}.

Therefore, in the thermodynamic potential $\Omega$, one  subtracts
from $\Omega$ the one for the ``ideal'' case
  \begin{eqnarray}
   \Omega(U=T=0)&=&\frac12\sum_{q}(\varepsilon({\bf q})-\mu-Jz_{0})\nonumber\\&=& \frac12 \sum_{q}(\varepsilon({\bf q})+\varphi ')\lab{21.1},
   \end{eqnarray}
   and deals only with the subtracted expression
   \begin{eqnarray}
   \Omega_{\rm ren}&=&\Omega(U,T)-\Omega(U=0,T=0)\nonumber\\&=&\frac12\sum_{q}({\cal E}({\bf q})-\varepsilon({\bf q})- \varphi ')+N_{s}\nu n_{\sbf 0}(\varphi '-\Delta)\nonumber\\
    {}&&+\displaystyle{\frac{N_{s}\Delta^2}{2U\sinh^2\theta}}-
 \displaystyle{\frac{N_{s}(\varphi '+\mu+Jz_{0})^2}{2U\cosh^2\theta}}\nonumber\\ \nonumber\\
{}&&+T\sum_{q}\ln(1-e^{-\beta {\cal E}({\bf q})})\lab{21},
     \end{eqnarray}
     where we have performed summation by Matsubara frequency by using formula
     \be
     \sum_{n=-1}^{\infty}\ln(\omega_{n}^2+a^2)=a\beta+2\ln(1-e^{-\beta a})+\textrm{divergent const.}
     \ee
  For brevity, we shall suppress writing down
the subtraction in $\Omega_{\rm ren}$.

In equilibrium, the thermodynamic potential reaches a minimum with respect to pa\-ra\-me\-ters $n_{\sbf 0}, \varphi '$
and $\Delta$. Thus we  minimize
$\Omega$ with respect to  $n_{\sbf 0}$
\be
 \frac{\partial\Omega}{\partial n_{\sbf 0}}=N_{s}\nu(\varphi '-\Delta)=0\label{22},
 \ee
and get
 \be
 \varphi '=\Delta\lab{22_1}.
 \ee
Inserting this into (\ref{18}) leads to the well-known Bogoliubov phonon dispersion
 \be
  {\cal E}({\bf q})=\sqrt{\varepsilon({\bf q})}\sqrt{\varepsilon({\bf q})+2\Delta}\lab{23},
  \ee
  which is linear in ${\bf q}$ for small momentum, thus respecting the Nambu-Goldstone theorem.

  Minimizing thermodynamic potential $\Omega$
with respect to $\Delta$ gives the equation:
  \be
  \Delta=U\sinh^2\theta \left[\nu n_{\sbf 0}+\dsfrac{\Delta }{N_{s}}
  \sum_{\sbf q}\dsfrac{c_{\sbf q}}{{\cal E}({\bf q})}\right]\lab{24},
  \ee
  where $  c_{\sbf q}$ stands for
  \be
  \ba
  c_{\sbf q}=\frac 1 2+f_{\beta}({\cal E}({\bf q}))
=\frac{1}{2}\coth {(\beta{\cal E}({\bf q})/2)},\\ \\ f_\beta(\omega)=1/{(e^{\beta\omega}-1)}.\lab{W}
\ea
\ee
 Minimizing $\Omega$ with respect to $\varphi '$, thereby
taking into account the relation $\partial{\cal E}(\bfq)/\partial\varphi '=(\varepsilon({\bf q})+\varphi ')/{\cal E}({\bf q})$,
gives the following equation:
  \be
  N_{s}\nu n_{\bf 0}+\sum_{q}\left[\frac{(\varepsilon(\bfq)+\varphi ')c_{\sbf q}}
  {{\cal E}(\bfq)}-\frac12 \right]-\frac{N_{s}(\varphi '+\mu+Jz_{0})}{U\cosh^2\theta}=0\lab{25}.
  \ee
This will serve to determine of  uncondensed  fraction $n_{\sbf u}$.

  \subsection{Normal and anomalous densities}

According to the general rules of statistical mechanics, the total
number of
  particles $N$ is conjugate to the chemical potential:
  \be
  N=-\left(\frac{\partial\Omega}{\partial\mu}\right)_{T,V}\nonumber.
  \ee
  Applying this to (\ref{21}) gives
  \be
  N=\dsfrac{N_{s}(\varphi '+\mu+Jz_{0})}{U\cosh^2\theta}\lab{26}.
  \ee
  Using (\ref{26}) in (\ref{25}),
we obtain
  \be
  N=N_{s}\nu n_{\sbf 0}+\sum_{q}\left[\frac{(\varepsilon(\bfq)+\varphi')c_{\sbf q}}
   {{\cal E}(\bfq)}-\frac12\right]\equiv N_{\sbf 0}+N_{\sbf u}
.\lab{27}
    \ee
Here $ N_{\sbf 0}$ is a total number of condensed atoms, and $n_{\sbf 0}=N_{\sbf 0}/N_{s}\nu$ is the
{\it condensate fraction}. The uncondensed atoms
have a fraction
    \be
     n_{\sbf u}=\frac{N_{\sbf u}}{N}=\frac{1}{\nu N_{s}}\sum_{q}\left[
      \frac{(\varepsilon(\bfq)+\varphi')c_{\sbf q}}{{\cal E}(\bfq)}-\frac12
       \right]\lab{28}.
        \ee
It satisfies the trivial
relation $n_{\sbf 0}+n_{\sbf u}=1$.

Note that, the term $-\frac12$ in the square bracket of (\ref{28}) is due to the
renormalization  procedure (\ref{21}), and guarantees that at $T=0$ all particles of the
ideal gas (which has $U=0$ and $ \Delta=0$) are condensed, so that
 $n_{\sbf u}(U=0,T=0)=0$.

When the U(1) gauge symmetry is broken, a Bose system is
characterized not only by the expectation values of the fluctuating
part of the $\psi$-field with the normal density $n_{\sbf u}=\langle
\tilde{\psi}^{\ast}\tilde{\psi}\rangle$, but also with anomalous
density, defined by \be
 \delta(x_{{\sbf i}},\tau,x_{j},\tau')=\langle\tilde{\psi}(x_{{\sbf i}},\tau)
  \tilde{\psi}(x_{j},\tau')\rangle\lab{29}.
   \ee
  Clearly, for homogenous system in the equilibrium, in particular,
  for periodic optical lattices without magnetic trap, $\delta$  does
  not depend on coordinates, i.e. $\delta(x_{{\sbf i}},\tau,x_{j},\tau')=$const
as  was emphasized in \ci{yukannals}. Omission  of the anomalous
averages makes all calculations not self-consistent, the dynamics
non-conserving, the thermodynamics incorrect. It ruins the order of
the phase transition and renders the system unstable. It was also
shown in \ci{yukannals} that a $\delta=0$ type of mean-field
approach referred in the literatures as Hartree-Fock-Popov (HFP)
approximations \ci{ourtriplon} leads to a discontinuity in the
magnetization curve of antiferromagnetic material with the  triplon
BEC. Thus we must always allow for  $\delta\neq 0$.

Let us calculate this expectation value from the formula
\bea
\delta&=&
\frac{1}{\nu}\langle\tilde{\psi}(x_{{\sbf i}},\tau)\tilde{\psi}(x_{{\sbf i}},\tau)\rangle\nonumber\\
 &=&\frac{1}{2\nu}[\langle\tilde{\psi}_{1}(x_{{\sbf i}},\tau)\tilde{\psi}_{1}(x_{{\sbf i}},\tau)\rangle
-\langle\tilde{\psi}_{2}(x_{{\sbf i}},\tau)\tilde{\psi}_{2}(x_{{\sbf i}},\tau)\rangle]
  \nonumber\\ & =&\frac{1}{2\nu}[G_{11}(0)-G_{22}(0)]\lab{30}.
    \eea
 In momentum space, the propagator can be rewritten as
  \begin{eqnarray}
   G_{ab}(\omega_{n},{\bf q})=\frac{1}{\omega_{n}^{2}+{\cal E}^{2}(\bfq)}
    \left(
\begin{array}{lr}
 {\varepsilon}({\bf q})+2\Delta & \omega_{n}\\
 -\omega_{n} &{\varepsilon}({\bf q})
\end{array}\right)\label{31},
  \end{eqnarray}
  where we used equations (\ref{8}), (\ref{12}), and (\ref{23}).
  Using in (\ref{30}) the equations (\ref{13_1}) and (\ref{31}), one obtains
  \be
   \ba
   \delta=\displaystyle{\frac{1}{2\nu N_{s}\beta}\sum_{n}\sum_{\bfq}
    \frac{2\Delta}{\omega_{n}^2+{\cal E}^{2}(\bfq)}}\\ \\
     \displaystyle{~= \frac{\Delta}{\nu N_{s}}
     \sum_{\bfq}\frac{c_{\sbf q}}{{\cal E}(\bfq)}= \frac{\Delta}
      {\nu N_{s}}\sum_{\bfq}\frac{1}{{\cal E}(\bfq)}\left(\frac12+ \frac{1}{e^{\beta{\cal E}(\bfq)}-1}\right)}\lab{32}.
       \ea
       \ee
       In terms of $\delta$, the $\Delta$-equation (\ref{24}) may be rewritten in the
        following compact form
       \be
        \Delta= U\nu(n_{\sbf 0}+\delta)\sinh^2\theta\lab{33},
         \ee
         with $n_{\sbf 0}=1-n_{\sbf u}$, and $n_{\sbf u}$ given by (\ref{28}).

It is well known that the Goldstone theorem for a dilute Bose
gas with a spontaneous broken
symmetry is equivalent
to the celebrated Hugenholtz-Pines theorem \ci{dickhoff},
according to which
 self-energy $\Sigma_{\rm cl}$ and the anomalous self-energy
 $\Delta_{\rm cl}$ satisfy
\be
 \Sigma_{\rm cl}-\Delta_{\rm cl}=\mu\lab{34}.
  \ee
  In the Appendix A we shall show that a similar equation
 holds for optical lattices:
  \be
  \Sigma_{\rm cl}-\Delta_{\rm cl}=\mu+Jz_{0}\lab{35},
  \ee
  with $\Sigma_{\rm cl}=\varphi _{0}\cosh\theta, ~\Delta_{\rm cl}=\Delta$.

  The only  parameter, that so far
 remains free in the initial action (\ref{1.5}), is $\theta$. It may be chosen
such that the quasiparticle energy ${\cal E}(\bfq)$ reduces,
 in the
one-loop approximation \cite{ourknr}, to the gapless
Bogoliubov dispersion
  \be
  {\cal E}(\bfq)_{\rm  one loop}=\sqrt{\varepsilon(\bfq)}\sqrt{\varepsilon(\bfq)+2U\nu}.
  \ee
Indeed, in this approximation we get from (\ref{33}) $\Delta\approx U\nu\sinh^2\theta $, and from (\ref{23})
 ${\cal E}(\bfq)\approx\sqrt{\varepsilon(\bfq)}\sqrt{\varepsilon(\bfq)+2U\nu\sinh^2\theta }$. This is the place where we fix the
$\theta$ to satisfy
 \be
 \sinh^2\theta=1,\qquad\cosh^2\theta=2\lab{33.1}
, \ee as was announced earlier.

Summarizing this section, we present  the
 full expression for $\Omega$:
 \bea
  \Omega&=&\frac12\sum_{\bfq}[{\cal E}(\bfq)-\varepsilon(\bfq)-\Delta]\nonumber\\ {}&& +
   \frac{N_{s}\Delta^{2}}{2U}
    -\frac{N_{s}(\Delta+\mu+Jz_{0})^2}{4U}\nonumber\\ \nonumber \\{}&&+
    T\sum_{\bfq}\ln(1-e^{-\beta{\cal E}(\bfq)})\lab{omloaf},
     \eea
 with
 \be
 \mu=2\nu U-\Delta-Jz_{0}\lab{mu}.
 \ee
 The last equation follows from (\ref{26}).
The self energy $\Delta$ in \re{omloaf} and \re{mu}  is defined through the following set  of
 nonlinear algebraic  equations:
 \be
 \ba
 \Delta=U\nu(n_0+\delta), \quad \quad n_0=1-n_{\sbf u},\\ \\
  n_{\sbf u}=\dsfrac{1}{\nu N_s}\dssum_{\bf q}\left [   \dsfrac{c_{\sbf q}(\varepsilon(\bfq)+\Delta)}{{\cal E}(\bfq)}-\dsfrac{1}{2}    \right ],\\
 \delta=\dsfrac{\Delta}{\nu N_{s}}
     \dssum_{\bfq}\frac{c_{\sbf q}}{{\cal E}(\bfq)},
      \ea
 \ee
 where $c_{\sbf q}$ is given in \re{W} and $U$, $J$, $\nu$,  $T$ are input parameters.

 \subsection{Symmetric phase}

 When $n_{\sbf 0}=0$,  the Hamiltonian (\ref{H11}) is symmetric under the transformation $\psi\to e^{i\alpha}\psi$ and  equation (\ref{22}) makes no sense. Then $\varphi '\neq\Delta$, and the energy spectrum has a gap with the dispersion
 \be
 {\cal E}(\bfq)=\sqrt{(\varepsilon(\bfq)+\varphi '-\Delta)
 (\varepsilon(\bfq)+\varphi '+\Delta)}\lab{41}.
 \ee
 The main equations in this regime with $T>T_{c}$ are
 \be
 \ba
 \Delta=U\nu\delta,~~~~~~
 \delta=\displaystyle{\frac{\Delta}{\nu N_{s}}\sum_{\bfq}\frac{c_{\bfq}}{{\cal E}(\bfq)}},\nonumber\\\\
 \nu=\displaystyle{\frac{1}{N_{s}}\sum_{\bfq}\left[\frac{(\varepsilon(\bfq)+\varphi ')c_{\bfq}}
  {{\cal E}(\bfq)}-\frac12\right]}\label{42}.
  \ea
  \ee
  The set of equations (\ref{42}) with the energy spectrum
(\ref{41}) may have a solution $\Delta\neq0,~ \varphi' >\Delta$,
  leading to an exotic state with no
condensate
 but with a finite anomalous density: $n_{\sbf 0}=0, \delta\neq0$. It was
shown in Ref. \ci{pra86} that this phase has a nonzero SF fraction.
The
 upper boundary of such a state was
 denoted by $T^{\ast}$, and was determined by solving
the
equations (\ref{42}) with
 $\Delta=0, ~\varphi '>0$.
Thus it was theoretically predicted that ultracold dilute atomic
gases posses a
 superfluid state at
$T_{c}<T\leq T^{\ast}$ without Bose condensation in the one-body
channel \ci{pra86}. However, up to date, such states have not been
observed experimentally.
  In Sect. IV we shall investigate
the possible existence of such a state for optical lattices, with a
negative outcome.

  \section{Variational Perturbation Theory
in optical lattices}
It   is  interesting to  compare our result with those of
Variational Perturbation Theory \ci{KS}. To lowest order, this is
equivalent to the HFB approximation used in the operator formalism
\ci{andersen}. To do this, let us formulate the HFB approximation
for optical lattices in the functional integral framework.

 Starting point is again (\ref{1.1}) in which we perform
the Bogoliubov shift (\ref{6}) and separate the action as follows
 \be
 \ba
 {\cal A} ={\cal A} _{(0)}+{\cal A} _{(1)}+{\cal A} _{(2)}+{\cal A} _{(3)}+{\cal A} _{(4)},
  \ea
  \ee
where
\begin{widetext}
 \be
 \ba
 {\cal A} _{(0)}=\beta N_{s}\nu n_{\sbf 0}[\frac{U}{2}\nu n_{\sbf 0}-\mu-Jz_{0}],\\ \nonumber\\
 {\cal A} _{(1)}=\displaystyle{\sqrt{\nu n_{\sbf 0}}[-\mu-Jz_{0}+U\nu n_{\sbf 0}]\int d\tau\sum_{{\sbf i}}
  (\tilde{\psi}(\bfx_{{\sbf i}},\tau)+\tilde{\psi}^{\ast}(\bfx_{{\sbf i}},\tau))},\nonumber\\
 {\cal A} _{(2)}=\displaystyle{\inttau d\tau\left\{ \sum_{{\sbf i}}\tilde{\psi}^{\ast}(\bfx_{{\sbf i}},\tau)
  [\partial_{\tau}-\mu]\tilde{\psi}(\bfx_{{\sbf i}},\tau)+\frac{U}{2}\nu
n_{\sbf 0}
\right.}\nonumber\\
  \left.~~\qquad\times \displaystyle{\sum_{{\sbf i}}}\left[\tilde{\psi}^{2}(\bfx_{{\sbf i}},\tau) +4\tilde{\psi}^{\ast}(\bfx_{{\sbf i}},\tau)
  \tilde{\psi}(\bfx_{{\sbf i}},\tau)+\tilde{\psi}^{\ast}(\bfx_{{\sbf i}},\tau)
  \tilde{\psi}^{\ast}(\bfx_{{\sbf i}},\tau)\right]-
  J\displaystyle{\sum_{\langle i,j\rangle}}\tilde{\psi}^{\ast}(\bfx_{{\sbf i}},\tau)
  \tilde{\psi}(\bfx_{j},\tau)\right\},\nonumber\\
 {\cal A} _{(3)}=\displaystyle{U \sqrt{\nu n_{\sbf 0}}\inttau
  d\tau\sum_{{\sbf i}}[\tilde{\psi}^{\ast}(\bfx_{{\sbf i}},\tau)\tilde{\psi}^{2}(\bfx_{{\sbf i}},\tau)
  +\tilde{\psi}^{\ast}(\bfx_{{\sbf i}},\tau)\tilde{\psi}^{\ast}(\bfx_{{\sbf i}},\tau) \tilde{\psi}(\bfx_{{\sbf i}},\tau)]},\nonumber\\
 {\cal A} _{(4)}=\displaystyle{\frac{U}{2} \inttau d\tau\sum_{{\sbf i}}[\tilde{\psi}^{\ast}(\bfx_{{\sbf i}},\tau)
  \tilde{\psi}(\bfx_{{\sbf i}},\tau) ]^{2}}\lab{43}.
  \ea
  \ee
After this we add and subtract following terms
  \be
   {\cal A} _{(\Sigma)}=\inttau d\tau\sum_{{\sbf i}}\left\{\Sigma_{\rm cl}\tilde{\psi}^{\ast}
   (\bfx_{{\sbf i}},\tau)\tilde{\psi}(\bfx_{{\sbf i}},\tau)+\frac12\Delta_{\rm cl}
   [\tilde{\psi}^{\ast}(\bfx_{{\sbf i}},\tau)\tilde{\psi}^{\ast}(\bfx_{{\sbf i}},\tau)+
   \tilde{\psi}(\bfx_{{\sbf i}},\tau)\tilde{\psi}(\bfx_{{\sbf i}},\tau)]\right\}\lab{44},
   \ee
    \end{widetext}
   with variational parameters $\Sigma_{\rm cl} $ and $\Delta_{\rm cl}$.
The subscripts cl emphasize that these are variational parameters
which, in contrast to the earlier fields $\varphi$ and $\Delta$, are
not meant to be functionally integrated.

Using again real and imaginary parts of the complex fields
$\tilde\psi ,~ \tilde\psi^\ast$ as in (\ref{1.12}),
 we
rewrite ${\cal A} $ as
\be
\ba
 {\cal A} ={\cal A} _{(0)}+{\cal A} _{\rm free}+{\cal A} _{\rm int},
  \ea
  \ee
where
\be
\ba
 {\cal A} _{\rm free}=\displaystyle{\frac12\inttau d\tau \sum_{{\sbf i}}\sum_{a,b=1,2}\psi_{a}
  (\bfx_{{\sbf i}},\tau)[i\varepsilon_{ab}\partial_{\tau}+Y_{a}\delta_{ab}]
  \psi_{b}(\bfx_{{\sbf i}},\tau)},\nonumber\\ \\
  {\cal A} _{\rm int}={\cal A} _{\rm int}^{(2)}+{\cal A} _{\rm int}^{(3)}+{\cal A} _{\rm int}^{(4)},\nonumber\\ \\
  {\cal A} _{\rm int}^{(2)}=\displaystyle{\frac12\inttau d\tau \sum_{{\sbf i}}\left\{\psi_{1}^{2}(\bfx_{{\sbf i}},\tau)[3U\nu n_{\sbf 0}-\Sigma_{\rm cl} -\Delta_{\rm cl}]\right.}\\
  \qquad+\left.\psi_{2}^{2}(\bfx_{{\sbf i}},\tau)[U\nu n_{\sbf 0}-\Sigma_{\rm cl} +\Delta_{\rm cl}]\right\},\nonumber\\
  {\cal A} _{\rm int}^{(3)}=\dsfrac{1}{2}U\sqrt{2\nu n_{\sbf 0}}\inttau d\tau \sum_{{\sbf i}}
   [\psi_{1}^{3}(\bfx_{{\sbf i}},\tau)+\psi_{1}(\bfx_{{\sbf i}},\tau)
   \psi_{2}^{2}(\bfx_{{\sbf i}},\tau)],\nonumber\\
  {\cal A} _{\rm int}^{(4)}=\dsfrac{1}{8}U\inttau d\tau \sum_{{\sbf i}}[\psi_{1}^{2}(\bfx_{{\sbf i}},\tau)+
  \psi_{2}^{2}(\bfx_{{\sbf i}},\tau)]^{2}\label{45},
  \ea
  \ee
  where
\be
\ba
Y_{1}=-\mu-Jz_{0}+\Sigma_{\rm cl} +\Delta_{\rm cl},\nonumber\\
Y_{2}=-\mu-Jz_{0}+\Sigma_{\rm cl} -\Delta_{\rm cl}\lab{46}.
\ea
\ee
 The free part of the action, ${\cal A} _{\rm free}$ in Eq. (\ref{45}), gives rise
to the propagator to be used in perturbation expansion. In the
momentum representation of the fields Eq. (\ref{9}), the propagator
is given by
  \begin{eqnarray}
   G(\omega_{n},{\bf q})=\frac{1}{\omega_{n}^{2}+{\cal E}^{2}(\bfq)}
    \left(
\begin{array}{lr}
 {\varepsilon}_{{\bf q}}+Y_{2} & \omega_{n}\\
 -\omega_{n} &{\varepsilon}_{{\bf q}}+Y_{1}
\end{array}\right)\label{47},
  \end{eqnarray}
 with ${\cal E}^{2}({\bf q})=({\varepsilon}_{{\bf q}}+Y_{1})({\varepsilon}_{{\bf q}}+Y_{2})$.
To lowest order, one obtains
 \be
 \Omega=-T\ln Z=-T\ln Z_{0}-T\ln Z_{\rm free}+T\langle{\cal A} _{\rm int}\rangle\lab{48},
 \ee
 where  $Z_{0}=e^{-{\cal A} _{(0)}},~ Z_{\rm free}=\dsint {\cal D}\psi_{1}{\cal D}\psi_{2}
e^{-{\cal A} _{\rm free}}=1/\sqrt{{\rm Det}\, G^{-1}},~\langle{\cal
A} _{\rm int}\rangle = \left\{\int{\cal D}\psi_{1}{\cal
D}\psi_{2}{\cal A} _{\rm int}e^{-{\cal A} _{\rm
free}}\right\}/Z_{\rm free}$.

 Now we evaluate
 \be
 \ba
 \displaystyle{\langle\psi_{a}^{2}(\bfx_{{\sbf i}},\tau)\rangle=G_{aa}(0)=\dsfrac{\sigma_{a}}{N_{s}}}, ~~~
 \langle\psi_{a}^{4}(\bfx_{{\sbf i}},\tau)\rangle=\dsfrac{3\sigma_{a}^2}{N^{2}_{s}}
, \nonumber\\\\
 \displaystyle{\langle\psi_{1}^{2}(\bfx_{{\sbf i}},\tau)\psi_{2}^{2}(\bfx_{{\sbf i}},\tau)\rangle=\dsfrac
 {\sigma_{1}\sigma_{2}}{N_{s}^2}}, \quad \langle{\cal A}_{\rm
 int}^{(3)}\rangle=0,
\lab{49}
 \ea
 \ee
 with
 \be
 \ba
\displaystyle{
\sigma_{1}=T\sum_{{\bf q},n}\frac{{\varepsilon}_{{\bf q}}+Y_{2}}{\omega_{n}^{2}+{\cal E}^{2}(\bfq)},
 \quad \sigma_{2}=T\sum_{{\bf q},n}\frac{{\varepsilon}_{{\bf q}}+Y_{1}}{\omega_{n}^{2}+{\cal E}^{2}(\bfq)},
}
 \lab{50}
 \ea
 \ee
and we find the following thermodynamic potential:
 \bea
 \Omega&=&N_{s}\nu n_{\sbf 0}\left(-\mu-Jz_{0}+\frac{U}{2}\nu n_{\sbf 0}\right)\nonumber\\
  {}&&+\frac{1}{2}\sum_{q}[{\cal E}({\bf q})-\varepsilon({\bf q})+\mu+Jz_{0}]\nonumber\\
  {}&&+T\sum_{{\bf q}}\ln(1-e^{-\beta{\cal E}({\bf q})})+\frac{U\nu}{8N}[3\sigma_{1}^{2}
  +3\sigma_{2}^{2}+2\sigma_{1}\sigma_{2}]\nonumber\\{}&&+\frac12 \sigma_{1}(3U\nu n_{\sbf 0}-Y_{1}-Jz_{0}-\mu)\nonumber\\{}&&
  +\frac12 \sigma_{2}(U\nu n_{\sbf 0}-Y_{2}-Jz_{0}-\mu)\lab{51},
  \eea
  where we have again subtracted $\Omega(T=0, ~U=0)$.

  The parameters $\Sigma_{\rm cl} $ and $\Delta_{\rm cl}$ are now
determined variationally by requiring that they minimize the
thermodynamic potential, i.e., we require
  $\partial\Omega/\partial\Sigma_{\rm cl}=0$ and $\partial\Omega/\partial\Delta_{\rm cl}=0$  \ci{HSp}, or equivalently
  \be
  \ba
  \dsfrac{\partial\Omega}{\partial Y_{1}}=0,\lab{52}~~~
  \dsfrac{\partial\Omega}{\partial Y_{2}}=0.\lab{53}
  \ea
  \ee
  These equations yield
  \be
  \ba
  Y_{1}=3U\nu n_{\sbf 0}-\mu-Jz_{0}+\dsfrac{U}{2N_{s}}(3\sigma_{1}+\sigma_{2}),\lab{54}\\
  \\
  Y_{2}=U\nu n_{\sbf 0}-\mu-Jz_{0}+\dsfrac{U}{2N_{s}}(\sigma_{1}+3\sigma_{2}).\lab{55}
  \ea
  \ee
The gaplessness of the energy spectrum is now imposed by hand. In fact,
by requiring the
relation  (\ref{35}), we get from (\ref{46}) $Y_{2}=0$ which leads
to the dispersion
  \be
  {\cal E}({\bf q})=\sqrt{\varepsilon({\bf q})}\sqrt{\varepsilon({\bf q})+2\Delta},\lab{55_1}
  \ee
  where $\Delta=Y_{1}/2$. This leads to the equations
  \be
  \ba
  \displaystyle{\Delta=U\nu n_{\sbf 0}+\frac{U}{2N_{s}}(\sigma_{1}-\sigma_{2})},\lab{56}\\
  \displaystyle{\mu+Jz_{0}=U\nu n_{\sbf 0}+\frac{U}{2N_{s}}(\sigma_{1}+3\sigma_{2})}.\lab{57}
  \ea
  \ee
  Here, we draw the reader's  attention
to the self-consistency of the HFB approximation
as far as the chemical potential is concerned. In fact, the stationary condition $\partial\Omega/\partial n_{\sbf 0}=0$ with $\Omega$ given by (\ref{51}) leads to the  following equation for $\mu$:
  \be
  \mu+Jz_{0}=U\nu n_{\sbf 0}+\frac{U}{2N_{s}}(3\sigma_{1}+\sigma_{2}),\lab{58}
  \ee
  which contradicts to $\mu$ of Eq.~(\ref{57}).

  To make the theory self-consistent, Yukalov and one of the authors \ci{yukalovkleinert}
  proposed to introduce two chemical potentials: namely, $\mu_{0}$, which corresponds to the Eq. (\ref{58}),
 and $\mu_{1}$ corresponding to Eq. (\ref{57}).  Being responsible for
 subsystem of condensed and uncondensed particles respectively they, naturally, coincide
 in the normal phase, when $Y_{1}=Y_{2}=0$. In the present  work,
however, we follow the standard procedure of identifying
$\mu$ in (\ref{57}) as a chemical potential from which we determine
the particle densities by differentiation of $\Omega$.

  \subsection{The fractions $n_{\sbf u}$ and $\delta$ in VPT}

  Applying the well-known relation $N=-\lfrac{\partial\Omega}{\partial\mu}$ to $\Omega$ in (\ref{51}) gives
  \be
  N=N_{s}\nu n_{\sbf 0}+\sum_{{\bf q}}\left[\frac{(\varepsilon({\bf q})+\Delta)c_{{\bf q}}}
  {{\cal E}({\bf q})}-\frac12\right]\equiv N_{\sbf 0}+N_{\sbf u},\lab{59}
  \ee
  and hence
  \be
  n_{\sbf u}=\frac{N_{\sbf u}}{N_{s}}=\frac{1}{\nu N_{s}}\sum_{{\bf q}}\left[\frac{(\varepsilon({\bf q})+\Delta)c_{{\bf q}}}
  {{\cal E}({\bf q})}-\frac12\right],\lab{60}
  \ee
  with the ${\cal E}({\bf q})$ is the Bogoliubovs dispersion given in (\ref{55_1}).\\
 For the anomalous density $\delta$ we obtain
  \bea
  \delta&=&\frac{1}{\nu}\langle\tilde{\psi}(\bfx_{{\sbf i}},\tau)\tilde{\psi}
(\bfx_{{\sbf i}},\tau)\rangle
  =\frac{1}{2N_{s}\beta\nu}[G_{11}(0)-G_{22}(0)]
\nonumber\\&=&\frac{(\sigma_{1}-\sigma_{2})}{2N_{s}\nu}
  =-\frac{\Delta}{\nu N_{s}}\sum_{{\bf q}} \frac{c_{{\bf q}}}{{\cal E}({\bf q})},\lab{61}
  \eea
  where we used Eqs. (\ref{47}) and (\ref{50}).

  Using now (\ref{61}) in (\ref{56}) gives the equation:
  \be
  \Delta=U\nu(n_{\sbf 0}+\delta),\lab{62}
  \ee
  which is formally the same as the one in before
 (\ref{33}) with (\ref{33.1}). The only difference between these two approximations is in the sign of anomalous density, which
 is, in general,
 $\delta>0$
in the collective quantum field theory and $\delta<0$ in HFB.

  Summarizing we collect here the main equations in both approximations:
  \bea
  \Delta&=&U\nu(n_{\sbf 0}+\delta),\lab{63}~~~
 n_{\sbf 0}=1-n_{\sbf u}\lab{64},\\
    \delta&=&\xi\frac{\Delta}{\nu N_{s}}\sum_{{\bf q}}\frac{c_{{\bf q}}}{{\cal E}({\bf q})},\lab{67}\\
 {\cal E}({\bf q})&=&\sqrt{\varepsilon({\bf q})}\sqrt{\varepsilon({\bf q})+2\Delta},\lab{66}\\
  c_{{\bf q}}&=&\frac12+\frac{1}{e^{\beta{\cal E}({\bf q})}-1},\lab{b8}\\
  \mu&=&2U\nu-\Delta-Jz_{0},\\
  \xi&=&\left\{\begin{array}{ll}
  -1, & \mbox{HFB}\\
  +1, & \mbox{Two Collective Quantum Fields and LOAF, }
   \lab{8999}
  \end{array}\right.
  \eea
where $n_{\sbf u}$ is given by (\ref{60}).

  Note that similar relations hold for atomic gases. A difference
occurs for the $T>T_{c}$ phase. There one may use replacements
listed in Appendix B.
  In fact, in the normal phase, $n_{\sbf 0}=0$, HFB theory gives
  \be
  \Delta=U\nu\delta=-\frac{\Delta}{\nu N_{s}}\sum_{{\bf q}}\frac{c_{{\bf q}}}{{\cal E}({\bf q})}\lab{70}.
  \ee
  Since the right-hand-side of this equation is positive, while
 the left-hand-side is negative, at least for optical lattices, Eq. (\ref{70})
 has the only solution  $\Delta=0$. This means that in
 the normal phase $n_{\sbf 0}=0$ and $\delta=0$ [see Eq. (\ref{67})] simultaneously. Therefore HFB
 theory does not predict a superfluid phase without a condensate,
 thus
 being
in contrast to the two-collective quantum field result
of
Cooper et.al. in Ref. \ci{pra86}
at the mean-field level.

  From above discussions it is easy to understand that VPT gives no shift
  in $T_{c}$  due to interaction. In fact, when $T\to T_{c}$, the condensed
  fraction $n_{\sbf 0}\to0$, and hence $\Delta\to0$. The expression for $n_{\sbf u}$, will
coincide  with that for the ideal gas,  i.e., Eq.~(\ref{60}) becomes
  \be
  \nu=\frac{1}{N_{s}}\sum_{{\bf q}}\frac{1}{e^{\beta\varepsilon({\bf q})}-1}\equiv
  \frac{1}{N_{s}}\sum_{{\bf q}}\frac{1}{e^{\varepsilon({\bf q})/T_{c}^{0}}-1},
  \ee
  which means that $T_{c}=T_{c}^{0}$ for HFB and, hence,  $\Delta T_{c}=T_{c}-T_{c}^{0}=0$.

  \section{Results and discussion}

  \subsection{Quantum phase transition in two-Collective Quantum Field
Theory and VPT
}

  First we discuss the existence of QPT in optical lattices for
two collective quantum fields at the mean field level and for the
HFB approximation. It has been shown that for dilute atomic Bose
gases Collective Quantum Field approximation does not predict QPT
\ci{pra84} while HFB does \ci{yukannals}. Below we show that in the
case of $3d$ optical lattices the situation is vice-versa. This can
be understood in the following way. Lets rewrite the main equation
at $T=0$ as:
  \be
  n_{\sbf 0}(\Delta)=\frac{\Delta}{U\nu}-\delta(\Delta)\lab{22_1}.
  \ee
  It is clear that for interacting system, $U\neq0$ and $\Delta\neq0$.
Since in the collective quantum field theory $\delta(\Delta)>0$, the
Eq. (\ref{22_1}) may have solution $n_{\sbf 0}(\Delta)=0$ with
$\Delta\neq0$ (see Table I). However, in HFB approximation
$\delta(\Delta)<0$ and $n_{\sbf 0}(\Delta)$ in (\ref{22_1}) may have
the only solution as $n_{\sbf 0}>0$ for $\Delta\neq0$. Note that in
the case of dilute atomic gases at $T=0$ \ci{ourchul}
  \be
  \ba
  \delta(\Delta)=-{8\rho\sqrt{\gamma/\pi}}<0 \quad \textrm{Two-Collective Quantum Field}\nonumber\\ \nonumber\\
  \delta(\Delta)=+{8\rho\sqrt{\gamma/\pi}}>0 \quad \textrm{HFB},\nonumber
  \ea\label{gasp}
  \ee
  with the dimensionless {\it gas parameter}
$\gamma=a_s^{3}\rho$ that characterizes the interaction strength of
the gas after renormalization. It is formed from the $s$-wave
scattering length $a_s$ and the particle density $\rho$. This sign
change is responsible for the
 dilute atomic gases has a QPT in the HFB approximation,
but not in the two-collective quantum field theory at the mean-field level.
{In Fig. 1, the  condensed  fraction $n_0$ as a
function of $u=U/J$ is presented for $\nu=1,2,3,4$. This may be
compared with $u_{\rm crit}=6(\sqrt{\nu}+\sqrt{\nu+1})^2$  given in
Gutzwiller's approximation.
  It is seen that although the two - collective quantum field theory predicts rather large
 value for $u_{\rm crit}$ (see Table I), it  gives desirable  second -  order phase
 transition.}

  \begin{figure}[h!]
\begin{center}
\leavevmode
\includegraphics[width=0.55\textwidth]{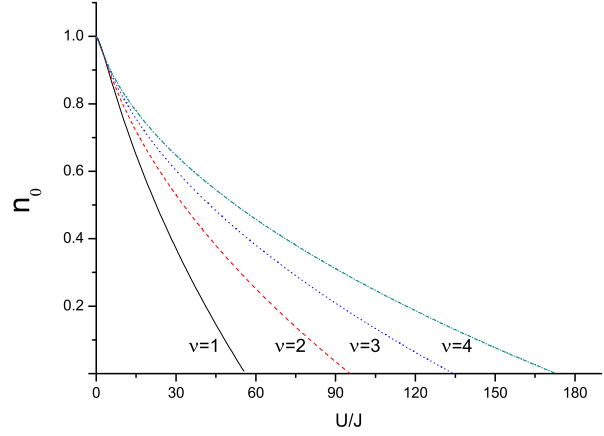}
\end{center}
\caption{(Color online) {The  condensed fraction $n_0$ at zero
temperature  as a function of}  $u=U/J$ for   various filling
factors, $\nu$. It is seen that $n_0$ goes to  zero smoothly and
vanishes at $u_{\rm crit}$. This may be compared with following
results by Gutzwiller's approximation: $u_{\rm crit}(\nu=1)=34.97$,
$u_{\rm crit}(\nu=2)=59.39$, $u_{\rm crit}(\nu=3)=83.56$, $u_{\rm
crit}(\nu=4)=107.66$. } \label{figQPT}
\end{figure}


  \subsection{Critical temperature $T_{c}^{0}$ for ideal cases}

  Before we
study the shift of $T_{c}$, let us
 estimate the critical temperature $T_{c}^{0}$ for the free optical lattice
with  $U=0$.
  Assuming $\Delta=0$ in Eq. (\ref{42}), we obtain the well-known formula
  \be
  \nu=\int_{0}^{1}dq_{1}dq_{2}dq_{3}\frac{1}{e^{\varepsilon_{q}/T_{c}^{0}}-1}\lab{23_1}.
  \ee
  Introducing dimensionless parameters $t_{c}^{0} =T_{0}^{c}/J,\quad
\hat{\varepsilon}_{\sbf q}=\varepsilon_{\bf
q}/2J=\sum_{\alpha=1}^{3}(1-\cos\pi q_{\alpha})$, we may rewrite
(\ref{23_1}) as
  \be
  \nu=\displaystyle{\int_{0}^{1}dq_{1}dq_{2}dq_{3}\frac{1}{e^{2\hat{
\varepsilon}_{\sbf q}/t_{c}^{0} }-1}}\lab{23_2}
  \ee
  which can be considered as a nonlinear equation for $t_{c}^{0} $ at a given filling factor $\nu$.
  Our numerical estimations for $t_{c}^{0} $ are given in Table \ref{tab1}. It is seen that
 for $\nu=1$, $T_{c}^{0}=5.6J$, which is in consistent  with other estimates
given in the references \ci{svistun, Yukalovobsor}.

\begin{table}[ tp]%
\caption {Critical parameters of Bose-Hubbard model vs. filling
factor $\nu$ in the two collective quantum field approach.
$u_{c}=(U/J)_{c}$ is given in the second row. The critical
temperatures of ideal optical lattices in $d=3$ are listed in units
$J$ in the third row. The forth row presents approximated values of
$t_{c}^{0} $ [see Eq. (\ref{23_3})]}
\begin{tabular}{|c|c|c|c|c|c|}
  \hline
  $\nu$                                      &   1   &    2  &     3 &    4  &  5      \\
  \hline
  $u_{c}=(U/J)_{c}$                        & 56.08 & 95.4 & 134.3 & 173  & 211.7 \\
  \hline
  $t_{c}^{0} =T_{c}^{0}/J$                 & 5.6   & 9.69  & 13.70 & 17.70  & 21.67  \\
  \hline
  $t_{c}^{0} $ in small $q$ approximation  & 5.06  & 10.07 & 15.2  & 20.25  & 25.32  \\
  \hline
  \end{tabular}
  \lab{tab1}
\end{table}

  Note that $T_{c}^{0}$
 can be approximated as $T_{c}^{0}=5.6J\nu^{0.825}$ in the range
 $\nu\in (1,5)$ including also non integer values.
  In the third row of Table I an approximated values of $t_{c}^{0} $ are presented.
  This approximation, say, spherical approximation at small momentum, is
  obtained by following replacements in (\ref{23_2}):
  \bea
  \int_{0}^{1}dq_{1}dq_{2}dq_{3}f({\bf q})\to\frac{\pi}{2}\int_{0}^{q_{d}}q^{2} dq f(q),\nonumber\\
  \hat{\varepsilon}_{\bfq}\to\frac{\pi^2}{2}{\bf q}^{2},\quad
  (e^{\varepsilon_{q}/T_{c}^{0}}-1)^{-1}
  \to\frac{T_{c}^{0}}{\varepsilon({\bf q})},\lab{23_3}
  \eea
  where the Debye momentum $q_{\rm D}$ defined by the equation:
  \be
  1=\int_{0}^1dq_{1}dq_{2}dq_{3}=\frac{\pi}{2}\int_{0}^{q_{d}}q^{2}dq,\lab{23_4}
  \ee
  equals to $q_{\rm d}=(6/\pi)^{1/3}\approx1.24$ for $d=3$. This gives $T_{c}^{0}/J=2\nu\pi(\pi/6)^{1/3}$.
  It is seen that this approximation works with roughly 10\% accuracy for $\nu\leq3$.

  \subsection{The shift in $T_{c}$ caused by the  interaction}

We are now prepared to estimate the shift $\Delta
T_{c}/T_{c}^{0}=(T_{c}-T_{c}^{0})/T_{c}^{0}$ analytically. Above we
have shown that the shift $\Delta T_{c}/T_{c}^{0}=0$ for VPT or
equivalently for HFB. For LOAF the integrals in the main equations
are dominated by small momenta. At $T\to T_{c}$ for $n_{\sbf 0}=0,
~n_{\sbf u}=1$ they are given by \bea
 \Delta&=&U\Delta\int_{0}^{1}dq_{1}dq_{2}dq_{3}\frac{f_{B}({\cal E}({\bf q}))}{{\cal E}({\bf q})},\lab{25_1}\\
 1&=&\frac{1}{\nu}\int_{0}^{1}dq_{1}dq_{2}dq_{3}\frac{(\varepsilon_{{\bf q}}+\Delta)}
 {{\cal E}({\bf q})}f_{B}({\cal E}({\bf q})),\lab{25_2}
 \eea
 with ${\cal E}({\bf q})=\displaystyle{\sqrt{\varepsilon_{{\bf q}}}\sqrt{\varepsilon_{{\bf q}}+2\Delta}}, \quad f_{B}({\cal E}({\bf q}))=1/(e^{\beta_{c}{\cal E}({\bf q})}-1),
 \quad \beta_{c}=1/T_{c}$.

 Note that in (\ref{25_1}) we may assume $\Delta\neq0$ and
divide both sides of (\ref{25_1}) by $\Delta$. The critical temperature of ideal gas
 $T_{c}^{0}$ is the solution of Eq.~(\ref{25_2}) with $\Delta=0$, i.e.,
 \be
 1=\dsfrac{1}{\nu}\dsint_{0}^{1}\frac{dq_{1}dq_{2}dq_{3}}{e^{\varepsilon_{{\bf q}}/T_{c}^{0}}
  -1}.\lab{25_3}
  \ee
  Now we introduce dimensionless variables:
  \be
  \ba
  \Delta=u^{2}\kappa^{2}T_{c}^{0},\quad T_{c}=T_{c}^{0}\alpha,\quad
T_{c}^{0}=Jt_{c}^{0} ,\\ \\
\varepsilon_{{\bf q}}=2J\hat{\varepsilon}_{{\bf q}},\quad
{\cal E}({\bf q})=2J\hat{{\cal E}}({\bf q}),
 \ea
  \ee
  with $\hat{\varepsilon}_{{\bf q}}=\sum_{\alpha}(1-\cos\pi{\bf q}_{\alpha}),\, \quad \hat{{\cal E}}({\bf q})=
\sqrt{\hat{\varepsilon}_{{\bf q}}}\sqrt{\hat{\varepsilon}_{{\bf q}}
+u^{2}\kappa^{2}t_{c}^{0} }, \quad {\Delta T_{c}}/{T_{c}^{0}}=\alpha-1$
   and $t_{c}^{0} $  are given in the third
row of Table I.

   The scaled equations can be rewritten as follows:
   \begin{eqnarray}
    {}&&\displaystyle{0=1-\frac{u}{2}\dsint_{0}^{1}\frac{f_{B}(\hat {{\cal E}}({\bf q}))dq_{1}dq_{2}dq_{3}}
     {\hat {{\cal E}}({\bf q})}}\lab{25_4},\\
     {}&&\displaystyle{0=1-\frac{1}{\nu}\dsint_{0}^{1} dq_{1}dq_{2}dq_{3} \frac{\hat {\varepsilon}_{{\bf q}}+u^{2}\kappa^{2}t_{c}^{0} /2}{\hat {{\cal E}}({\bf q})}
     f_{B}(\hat {{\cal E}}({\bf q}))}\lab{25_5},
  \end{eqnarray}
with $f_{B}(\hat {{\cal E}}({\bf q}))=1/(e^{2\hat {{\cal E}}({\bf q})/\alpha t_{c}^{0} }-1)$.\\

Bearing in mind (\ref{25_3}), we may rewrite (\ref{25_5}) as \be
\dsint_{0}^{1} dq_{1}dq_{2}dq_{3}\left\{\frac{1}{e^{2\hat
{{\cal E}}({\bf q})/t_{c}^{0} }-1}
 -\frac{\hat {\varepsilon}_{{\bf q}}+u^{2}\kappa^{2}t_{c}^{0} /2}{\hat {{\cal E}}({\bf q})
 (e^{2\hat {{\cal E}}({\bf q})/\alpha t_{c}^{0} }-1)}\right\}=0\lab{26_1}.
\ee The nonlinear equations (\ref{25_4}) and (\ref{26_1}) should be
solved with respect to $\kappa$ and $\alpha$ with given numbers
$u=U/J$ and $t_{c}^{0} $. To do this we make  replacements
(\ref{23_3}). Then  Eqs.  (\ref{25_4}) and (\ref{26_1}) can be
rewritten as
\bea \displaystyle{1-\frac{u\alpha t_{c}^{0}
}{4\sqrt{2}\pi^{2}}\int_{0}^{\varepsilon_{\rm D}}
\frac{d\varepsilon}{\sqrt{\varepsilon}(\varepsilon+u^{2}\kappa^{2}t_{c}^{0} )}=0}\lab{26_3},\\
\\
\displaystyle{\int_{0}^{{\varepsilon_{\rm D}}}\frac{d\varepsilon}{\sqrt{\varepsilon}}\left
\{1-\frac{\alpha(\varepsilon+u^{2} \kappa^{2}t_{c}^{0} /2)}
{\varepsilon+u^{2}\kappa^{2}t_{c}^{0} }\right\}=0}\lab{26_4},
\eea
where ${\varepsilon_{\rm D}}=\pi^{2}q_{\rm D}^{2}/2=(\pi^{2}/2)(6/\pi)^{2/3}$

The integrals in (\ref{26_3}) and (\ref{26_4}) are easily done and
yield
\bea
    0&=&\sqrt{2}(6\pi^{2})^{1/3}(1-\alpha)+u\alpha \kappa\sqrt{t_{c}^{0} }\arctan\tilde{\theta}\lab{26_6a}\\
    \\
     0&=&4\pi^{2}\kappa-\sqrt{2}
     \alpha\sqrt{t_{c}^{0} }
     \arctan\tilde{\theta},\lab{26_6}
  \eea
  where $\tilde{\theta}=\sqrt{2}(6\pi^{2})^{1/3}/(2\kappa u\sqrt{t_{c}^{0}
  })$. Excluding $\alpha$ from (\ref{26_6}) and inserting it to (\ref{26_6a})
   gives
  \bea
     {}\alpha&=&\displaystyle{\frac{2\sqrt{2}\pi^{2}\kappa}{\sqrt{t_{c}^{0} }\arctan{\tilde{\theta}}}}\lab{26_7},\\
0&=&
     {}4\kappa\pi^{8/3}6^{1/3}-\sqrt{2t_{c}^{0} }[(6\pi^{2})^{1/3}+2u \kappa^{2}\pi^{2}]\arctan{\tilde{\theta}}.\lab{26_8}\nonumber\\
     \eea

     Now we consider separately two regimes:

     {\bf a) Weak interacting regime.} Expanding (\ref{26_7}) and (\ref{26_8}) in linear order by $u$ we get
     \bea
       {}\alpha&=& \displaystyle{\frac{4\pi^{2}\kappa\sqrt{2}}{\sqrt{t_{c}^{0} }}+\frac{8\kappa^{2}u}
         {3}\left(\frac{6}{\pi}\right)^{2/3}}\lab{27_1},\\
         {}\kappa&=&\displaystyle{\frac{\sqrt{2t_{c}^{0} }}{8\pi}}.\lab{27_2}
         \eea
   Now inserting $\kappa$ into (\ref{27_1}) we finally obtain
   \be
   \alpha=1+\frac{ut_{c}^{0} }{12}
    \left(\frac{6}{\pi^{4}}\right)^{2/3}+O(u^{2})\lab{27_3},
   \ee
and hence \be \frac{\Delta T}{T_{c}^{0}}=\alpha-1=\frac{ut_{c}^{0}
}{12}
 \left(\frac{6}{\pi^{4}}\right)^{2/3}+O(u^{2}),\lab{27_4}
  \ee
  which means that for small coupling constant, i.e. $(U/J)<1$, the shift is positive and increases with $U/J$.

  {\bf b) Strong interacting regime.} In this region, $\Delta/u^{2}$ and hence, $\kappa$ is small, so
 we may  use a
linear approximation in $\kappa$ in Eqs. (\ref{26_7}), (\ref{26_8})
   \bea
       {} \alpha&=&\displaystyle{\frac{4\pi^{2}\kappa\sqrt{2}}{\sqrt{t_{c}^{0} }}}\lab{27_5},\\
0&=&         \displaystyle{\frac{\sqrt{2t_{c}^{0} }(6\pi^{5})^{1/3}}{2}-\kappa[2ut_{c}^{0}
        +4(6\pi^{8})^{1/3}]}\lab{27_6}.
         \eea
This leads to following equation \be \alpha=\frac{2
\pi^{8/3}6^{1/3}}{ut_{c}^{0}
+2\pi^{8/3}6^{1/3}}=\frac{T_{c}}{T_{c}^{0}}, \ee from which one may
conclude that $T_{c}$ decreases with increasing $u$, i.e. \be
\frac{\Delta T_{c}}{T_{c}^{0}}=\alpha-1=-\frac{ut_{c}^{0} }
{ut_{c}^{0}+2\pi^{8/3}6^{1/3}}<0. \ee Thus, our analytical estimate
shows that the critical temperature $T_{c}$ as a function of the
coupling constant $U$, i.e. the function $T(u)$ first increases and
then decreases with increasing $u$ for optical lattices. The
suppression of $T_{c}$ at large coupling  constant is in agreement
with experimental measurements \ci{trotzky} .

In Fig. 2 we present $T_{c}$ (in unit of $J$) vs. $u$ for $\nu=1$.
The solid line correspond to the exact numerical calculation, i.e., the
 numerical solutions of Eqs.  (\ref{25_1}), (\ref{25_2}).
 The experimental points (circles) are taken from \ci{trotzky},
 solid diamonds are from Monte-Carlo calculations  taken from ref.\ci{svistun}
 The suppression of $T_{c}$ at large coupling  constant is found for integer  $\nu\geq1$ also, as it is
seen in Fig. 3.

\begin{figure}[h!]
\begin{center}
\leavevmode
\includegraphics[width=0.55\textwidth]{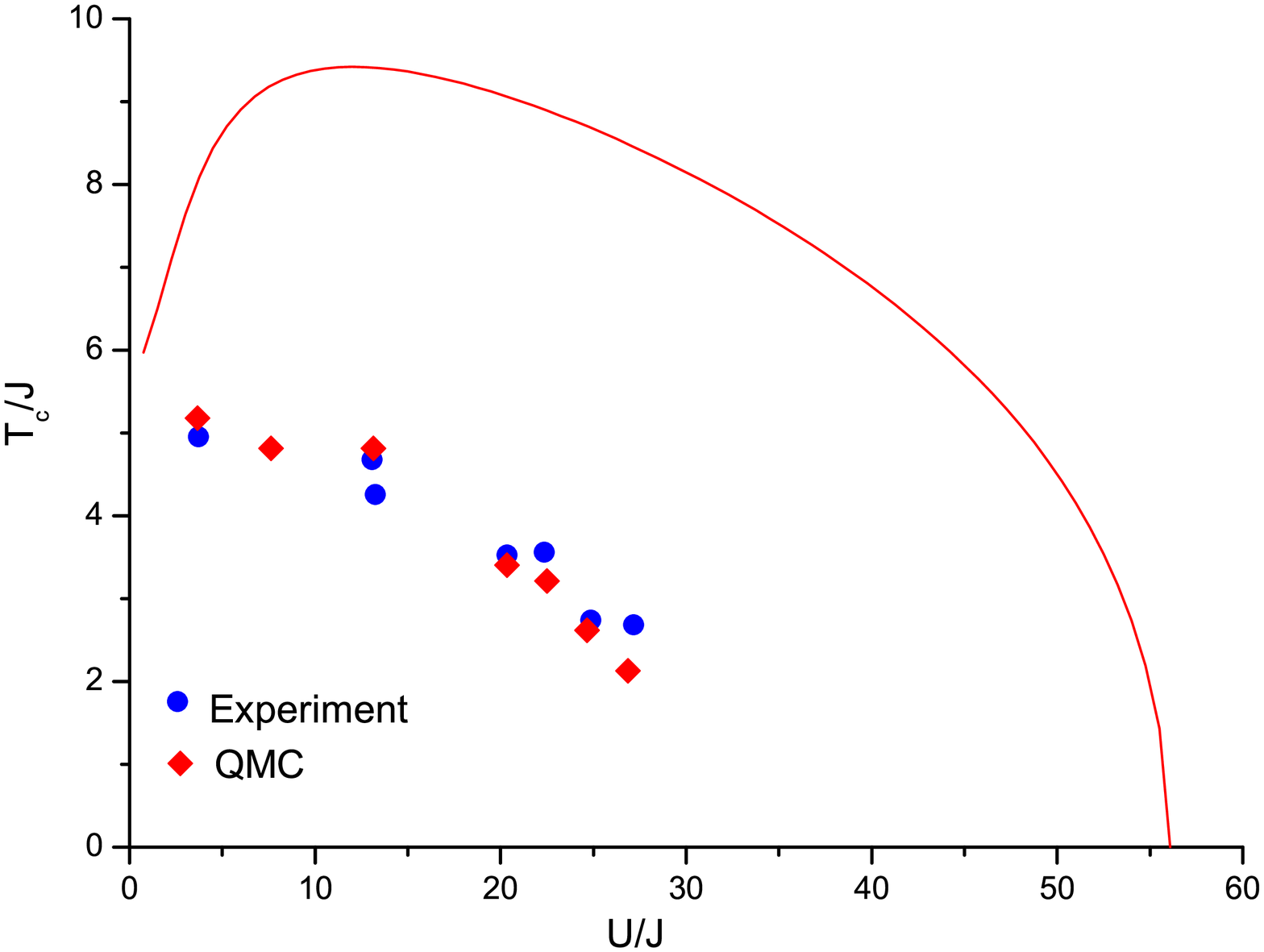}
\end{center}
\caption{(Color online) Behavior of $T_{c}$ (in units J) as a
function of $U/J$ in the saddle-point approximation to the
two-collective field theory for $\nu=1$. The circles show
experimental values given in \ci{trotzky}, solid diamonds are from
Monte-Carlo calculations of Ref.~\ci{svistun}. Note the initial rise
that was found also in atomic gases in Ref.~\cite{Kl5l}.}
\label{fig1}
\end{figure}

\begin{figure}[h!]
\begin{center}
\leavevmode
\includegraphics[width=0.55\textwidth]{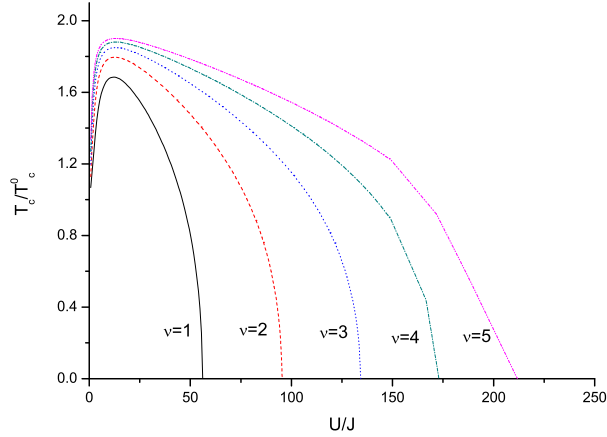}
\end{center}
\caption{(Color online) The same curves as in Fig. 1 but for $\nu=1,
2, 3, 4, 5$.} \label{fig1}
\end{figure}

In Fig. 4 we present the critical values of the self energy $\Delta_{c}=\Delta(T=T_{c})$ in units $J$ vs. $(U/J)$.
\begin{figure}[h!]
\begin{center}
\leavevmode
\includegraphics[width=0.55\textwidth]{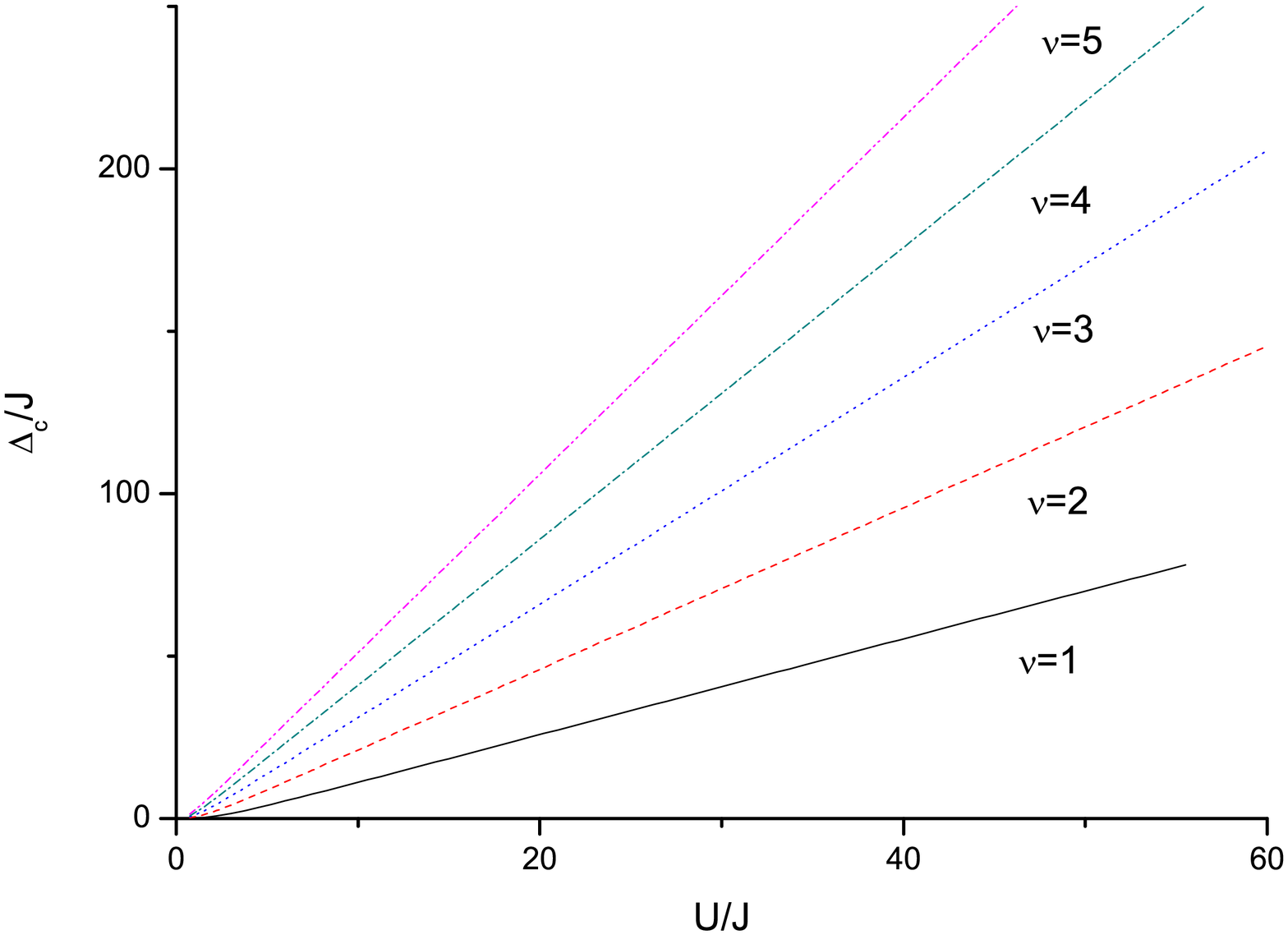}
\end{center}
\caption{(Color online) The critical $\Delta_{c}$ vs. $u$ for various filling factors $\nu$.}
\label{fig1}
\end{figure}
Observe that when $J$ is fixed, $\Delta_{c}$ increases with increasing $u$ and $\nu$.
On the other hand we observed that $\Delta_{c}$ in units $T_{c}^{0}(\nu)$,
i.e. $\Delta_{c}(T=T_{c})/T_{c}^{0}(\nu)$ vs. $u$ is almost independent on $\nu$,
e.g., $(\Delta_{c}/T_{c}^{0})|_{\nu=1}=7.656$ and $(\Delta_{c}/T_{c}^{0})|_{\nu=4}=7.780$ at $u=42.0$.

Now we consider the behavior of $\Delta$ for $T>T_{c}$.
It was suggested by
 Cooper et.al. \ci{pra86} that in the temperature
range $T\in (T_{c},T^{*})$ there exists
a
$U(1)$-symmetric phase with $n_{\sbf 0}=0$  but $\sigman\neq0$.
This would imply the existence of
a superfluid state
without a condensate.
 However, by solving
 (\ref{42}) for $\Delta$ and $\chin '$,
we could not find,  for optical lattices, any solution with
 $\Delta\neq0, \chin '\neq0$. Instead,  the equations for $T>T_{c}$,
 have a solution with $\Delta=0, \chin '=2\Un\nu-Jz_{0}-\mu$.
 In this normal state with $\sigman=0$,
the filling factor that characterizes the
particle density, is determined by the well-known equation
 \be \ba
\displaystyle
\nu=\frac{1}{N_{s}}\sum_{\sbf q}\frac{1}{e^{\beta(\varepsilon_{\sbf q}-2U\nu-Jz_{0}-\mu)}-1}

\nonumber\\
\quad=\displaystyle{\int_{0}^{1}dq_{1}dq_{2}dq_{3}
\frac{1}{e^{\beta(\varepsilon_{\sbf q}-2U\nu-Jz_{0}-\mu)}-1}}\lab{30_1},
\ea \ee
with the bare dispersion $\varepsilon_{\sbf q}=2J\sum_{\alpha=1}^{3}(1-\cos\pi q_{\alpha})$.\\
The chemical potential of interacting bosons in $T>T_{c}$ may be evaluated self consistently
from Eq. (\ref{30_1})  with input parameters $\nu, ~J,~ U$, and $T$, or given by an external
field (pumping) as in the case of triplons \ci{ourtriplon, ourdisorder}.

\section{Conclusion}
In this paper we have developed a Collective Quantum Field Theory
and a Variational Perturbation Theories for $d=3$ optical lattices
at very low temperatures. Both approximations satisfy
Hugenholtz-Pines theorem.
 We have shown that,
a two-Collective Quantum Field treatment in the saddle point
approximation predicts a second - order Quantum Phase transition,
that is missed in the VPT \ci{yukalovcond}. Unfortunately, the
predicted critical value of $(U/J)_{c}$ e.g. for $\nu=1$ is nearly
twice as large  as the experimental one.
 {Note that the main equations of  the previously
mentioned
approximation LOAF
\ci{pra83}  (recall  page \pageref{LOA})
and VPT are formally the same. The difference is in the
sign of the anomalous density $\delta$, as it is  seen from
equations \re{63}--\re{8999}}. We obtained analytical estimation
for the shift of critical temperature $T_{c}$ due to the point
interaction both in the weak and strong interaction regimes. It is
zero for VPT, while it has a nontrivial dependence on the coupling
strength
 $(\Un/J)$ in the Collective Quantum Field treatment
as well as in the LOAF approximation.
{The  general  behavior   of the phase diagram
compares
qualitatively well with existing experimental and ab initio
quantum Monte Carlo results.
 The similar   behavior e.g. suppression of the critical
temperature at large gas parameter for homogenous interacting Bose
gases have also found in Path - Integral - Monte - Carlo simulations
\ci{prokofev}}.
As to the dependence of the critical temperature on the filling
factor, $T_{c}/T_{c}^{0}$ increases with increasing $\nu$ at fixed
$U/J$.
 {From figures Fig. 1 and Fig. 2 one may conclude that in
order to describe the phase transitions in optical lattices more
accurately, the present theory should be extended beyond the saddle
point approximation used in Eq. \re{16}}, or in the spirit of B-DMFT
\ci{anders}.
{ We have found no exotic superfluid state with
finite anomalous density but zero condensate. Therefore, the
temperatures
 $T^{*}$ and $T_{c}$
introduced by
 Cooper et.al. \ci{pra86}
coincide. The system is in superfluid state for $0\leq T\leq T_{c}$,
and in normal state for $T>T_{c}$. It is natural that the
condensation will always be present in the one-body channel (see Eq.
(\ref{6})).}


 \section*{Acknowledgments}

 The work is supported in part by DAAD and Uzbek Science Foundation. We  are indebted to
 Fred  Cooper   for  useful discussions.

\newpage

\appendix
\section*{Appendix A}

 Below we derive Hugenholtz-Pines theorem
 \be
 \Sigma_{\rm cl}-\Delta_{\rm cl}=\mu+Jz_{0}\lab{a1},
 \ee
of Sect. II 
 for optical lattices.
 The normal $\Sigma_{\rm cl}$,and anomalous $\Delta_{\rm cl}$ self-energies in (\ref{a1})
 correspond to the normal $G_{\rm n}(r,r')=\langle T_{\tau}
\tilde{\psi}(r)\tilde{\psi}^{+}(r')\rangle$
 and anomalous $G_{\rm an}(r,r')=-\langle T_{\tau}\tilde{\psi}(r)\tilde{\psi}(r')\rangle$ Green
 functions respectively.
 In the Cartesian parametrization of the
quantum field (\ref{1.12})  we have:
 \bea
 \Sigma_{\rm cl}=\frac12[\Pi_{11}+\Pi_{22}],\lab{a2}\\
  \Delta_{\rm cl}=\frac12[\Pi_{22}-\Pi_{11}],\lab{a3}
  \eea
  where $\Pi_{ab}$ are defined by Dyson-Beliaev equations \ci{shi}:
  \be
  (\hat{G}^{-1})_{ab}-(\hat{G}^{-1}_{0})_{ab}=\Pi_{ab},\lab{a4}
  \ee
and the Green function $\hat{G}_{0}$ corresponds to the noninteracting situation
    \begin{eqnarray}
   G_{0}^{-1}(\omega_{n},{\bf q})=
    \left(
\begin{array}{lr}
 {\varepsilon}({\bf q})-\mu-Jz_{0} & -\omega_{n}\\
 \omega_{n} &{\varepsilon}({\bf q})-\mu-Jz_{0}
\end{array}\right).\label{a5}
  \end{eqnarray}
The interacting Green function $\hat{G}^{-1}$ is defined in Eq.(\ref{12}). Using (\ref{8}),
(\ref{17_1}), (\ref{12}), (\ref{a5}) in (\ref{a4}) gives:
 \be
 \ba
 \Pi_{11}=X_{1}+\mu=\cosh\theta\varphi_{0}-\Delta,\\
 \Pi_{22}=X_{2}+\mu=\cosh\theta\varphi_{0}+\Delta,\\
 \Pi_{12}=\Pi_{21}=0.\lab{a6}
 \ea
 \ee
 Inserting (\ref{a6}) into (\ref{a2}) and (\ref{a3}) one derives
 \be
 \ba
 \Sigma_{\rm cl}=\varphi_{0}\cosh\theta,\\
 \Delta_{\rm cl}=\Delta.\lab{a7}
 \ea
 \ee
 and hence
 \be
 \Sigma_{\rm cl}-\Delta_{\rm cl}=\varphi_{0}\cosh\theta-\Delta=\varphi'+\mu+Jz_{0}-\Delta,
 \lab{a8}
 \ee
 where we have used Eq.(\ref{17_1}).
 As it has been shown in Sect. II, in the condensed phase
 $\varphi'=\Delta$ and Eq.(\ref{a8})
becomes equivalent to the Hugenholtz-Pines theorem, i.e. to Eq.(\ref{a1}).

 The relation (\ref{a1}) in HFB approximation can be proved in a similar way.

 \newpage

 \section*{Appendix B}

 Here we present formal equivalence between Bose-Hubbard Hamiltonian (\ref{H11})
 in Wannier representation and standard Hamiltonian for homogeneous  dilute atomic gases
 \be
 H=\int d\bfr\Psi^{\dag}(\bfr)\left[-\frac{\vec{
 \nabla}^{2}}{2m}-\mu\right]\Psi
 (\bfr)+\frac{g}{2}\int d\bfr
 [\Psi^{\dag}(\bfr)\Psi(\bfr)]^{2}\lab{b1},
\ee
 where $g$ is the constant of contact interatomic interaction.
Using the replacements listed in Table II we obtain for $\Omega$
and the extremality equations in dilute atomic gases versus optical lattices
the relevant quantitied as derived in
Sections II and III. Of course,  an
appropriate renormalization procedure is implied
in dilute atomic gases.

\begin{widetext}

\begin{table}[h]\caption{Formal similarity between Hamiltonians (\ref{H11}) and (\ref{b1})}

\scriptsize{
\begin{tabular}{|l|c|c|c|}
  \hline
  Quantity               &  Homogeneous atomic gases   &    $3D$ Bose-Hubbard model   &     Comment       \\
  \hline
  Volume                  & $V$           & $N_{s}$ÿ                   & $N_{s}$-- number of sites \\
\hline
  Density                & $\rho=N/V$ÿ     & $\nu=N/N_{s}$                & $\nu$-- filling factor \\
\hline
  Bare dispersion       & $\varepsilon({\bf q})={\bf q}^{2}/2m$ & $\displaystyle{\varepsilon({\bf q})=2J\sum_{\alpha=1}^{3}(1-\cos\pi q_{\alpha})}$ & No additional magnetic trap  \\
  \hline
  Chemical potential       &  $\mu$         &  $\mu+Jz_{0}$                 &  $N=-\left(\dsfrac{\partial\Omega}{\partial\mu}\right)_{T}$\\
  \hline
  Momentum summation        & $\dsfrac{1}{V}\displaystyle{\sum_{q}f(\varepsilon({\bf q}))=\dsfrac{1}{2\pi^2}\int_{0}^{\infty}q^2dqf(\varepsilon(q))}$ &
                              $\dsfrac{1}{N_s}\displaystyle{\sum_{q}f(\varepsilon({\bf q}))=\int_{0}^{1}dq_{1}dq_{2}dq_{3}f(\varepsilon(q))}$ & $d=3$  \\
  \hline
  Normalization of densities&  $\rho_{0}+\rho_{1}=\rho$ &  $n_{0}+n_{1}=1$& In  the condensed phase.\\
                             &                          &                 &   No disorder.\\
  \hline
  \end{tabular}\lab{tab2}
  }
\end{table}
\end{widetext}

\bb{99}
\bi{Morchrev} O. Morsch and M. Oberthaler Rev. Mod. Phys. {\bf 78},
179 (2006).
 \bi{rous2003}  R. Raussendorf, D.E. Browne, and H.J. Briegel, Phys. Rev. A {\bf 68}, 022312 (2003).
\bi{svistun}B. Capogrosso-Sansone, N.V. Prokofev, and B.V.
Svistunov, Phys. Rev. {\bf B} 75, 134302 (2007). \bi{trotzky}
Trotzky, L. Pollet, F. Gerbier, U. Schnorrberger, I. Bloch, N.V.
Prokofev, B. Svistunov and M. Troyer, Nature Phys. {\bf 6}, 998
(2010).

\bi{stoofbook} H.T.C. Stoof, K.B. Gubbels, and D.B.M. Dickerscheid,
{\it Ultracold Quantum Fields} (Springer, 2009).

\bi{lewenstein} M. Lewenstein, A. Sanpera, and V. Ahufinger, {\it
Ultracold atoms in optical lattices: Simulating quantum many-body
systems} (Oxford University Press, 2012).

\bi{ueda}  M. Ueda, {\it Fundamentals and new frontiers of Bose-
Einstein condensation} (World Scientific, Singapore, 2010).

\bi{freericks} J.K. Freericks, H.R. Krishnamurthy, Yasuyuki Kato,
Naoki Kawashima, and Nandini Trivedi, Phys. Rev. A {\bf 79},
053631--1-22 (2009).

\bi{Pelster}F.E.A. dos Santos and A. Pelster, Phys. Rev. A {\bf 79},
013614 (2009).

 \bi{dutta} A. Dutta, C. Trefzger, and K.
Sengupta, arXiv:1111.5085v3 (2012).

\bi{luhman}D.-S. L\"uhmann, Phys. Rev. A {\bf 87}, 043619 (2013).

\bi{amico} L. Amico and V. Penna, Phys. Rev. Lett. {\bf80},
2189-2192 (1998).


\bi{vezzani} P. Buonsante and A. Vezzani, Phys. Rev. A {\bf 70},
033608 (2004). \bi{BV} K. Byczuk  and D. Vollhardt   Phys. Rev. B
{\bf 77}, 235106 (2008). \bi{anders} P. Anders  et al., New J. Phys.
{\bf 13}, 075013 (2011) \bi{dupius} A. Rancon and N. Dupuis Phys.
Rev. A {\bf 86}, 043624 (2012) \bi{Stoofmakola}D. van Oosten, P. van
der Straten, and H.T.C. Stoof Phys. Rev. A {\bf 63}, 053601 (2001).
\bi{ourknr} H. Kleinert, Z. Narzikulov, and Abdulla Rakhimov, Phys.
Rev. A {\bf 85}, 063602 (2012). \bi{pra83} F. Cooper, B. Mihaila,
J.F. Dawson, C.C. Chien, and E. Timmermans, Phys. Rev. A {\bf 83},
053622 (2011).

\bi{pra84} B. Mihaila, F. Cooper, J.F. Dawson, C.C. Chien, and E.
Timmermans, Phys. Rev. A {\bf 84}, 023603 (2011).

\bi{baym}  G. Baym, J.-P.Blaizot, M.Holzmann, F. Laloe, and
D.Vautherin, Phys. Rev. Lett. {\bf 83}, 1703 (1999).
 \bi{Kl5l} H. Kleinert,
Mod. Phys. Lett. B {\bf17}, 1011 (2003) (klrt.de/320).

\bi{pra86} J.F. Dawson, B. Mihaila, and F. Cooper, Phys. Rev. A {\bf
86}, 013603 (2012);\\ J.F. Dawson, F. Cooper, C.-C. Chien, and B.
Mihaila, Phys. Rev. A {\bf 88}, 023607 (2013).
 \bi{ourtriplon} Abdulla Rakhimov,
S. Mardonov, and E. Ya. Sherman, Annals of Phys. {\bf 326}, 2499
(2011). \bi{ourdisorder} Abdulla Rakhimov, S. Mardonov, E. Ya.
Sherman, and A. Schilling, New J. Phys. {\bf 14}, 113010 (2012).
\bi{GFCM} For field theories on a lattice see
 H. Kleinert, {\it Gauge Fields in Condensed Matter\/},
     Vol.~I \,\,  Superflow and Vortex Lines,
     World Scientific, Singapore 1989 (klnrt.de/b1).

\bi{CQFp}H. Kleinert,
Fortschr. Phys. {\bf 26}, 565 (1978) (klnrt.de/55).

\bi{CQFb}H. Kleinert,
{\it Collective Classical an Quantum Fields}, World Scientific, Singapore,
2013
(klnrt.de/b7).

\bibitem{HEA}
H. Kleinert, Fortschr. Phys. {\bf 30}, 187 (1982).

\bibitem{FKL}
R.P. Feynman and H. Kleinert, Phys. Rev. A {\bf34}, 5080 (1986).

\bibitem{KS}

H. Kleinert and Schulte-Frohlinde,
{\it Critical Properties of $\Phi^4$-Theories},
World Scientific, Singapore 2001
 ({\tt klnrt.de/b8}).

\bibitem{KLVPT}
H. Kleinert, EJTP {\bf 8}, 15 (2011) (klnrt.de/387).

 \bi{Yukalovobsor} V. I. Yukalov, Laser
Physics {\bf 19}, 1 (2009).

\bi{HK} H. Kleinert,
EJTP {\bf8}, 25 (2011) (klnrt.de/391).

 \bi{danshita}I. Danshita and
P. Naidon, Phys. Rev. A {\bf 79}, 043601  (2009).

\bi{REM1}
See Section 4.3 in Ref.~\cite{PI}.

\bi{PI}
H. Kleinert,
 {\it Path Integrals in Quantum Mechanics, Statistics and Polymer Physics,\/}
  World Scientific Publishing Co., Singapore 1995
(klnrt.de/b5). See the discussion in Subect. 2.15.2.


\bi{haugset} T. Haugset, H. Haugerud, and F. Ravndal,
    Ann. Phys. {\bf 27}, 266 (1998).
\bi{yukannals} V. I. Yukalov, Ann. Phys. {\bf 323}, 461 (2008).
\bi{dickhoff} W. H. Dickhoff and D. Van Neck, {\it Many-Body Theory
Exposed} (World Scientific, 2005).

\bi{andersen}  J. O. Andersen, Rev. Mod. Phys. {\bf 76}, 599 (2004).

\bi{HSp}H. Kleinert, EJTP {\bf 8}, 57 (2011) (klnrt.de/391).

\bi{yukalovkleinert} V. I. Yukalov and H. Kleinert, Phys. Rev. A
{\bf73}, 063612 (2006).

\bi{ourchul} A. Rakhimov, Chul Koo Kim , Sang-Hoon Kim, Jae-Hyung
Yee, Phys. Rev. A {\bf77},  033626 (2008).

\bi{yukalovcond}V. I. Yukalov, Condensed Matter Physics, 16, 23002
(2013).

\bi{prokofev} S. Pilati, S. Giorgini, and N. Prokof'ev, Phys. Rev.
Lett. {\bf 100}, 140405 (2008).

 \bi{shi} H. Shi and A. Griffin,
Phys.Rep. 304, 1 (1998).

 \eb

\edc